\documentclass{aa}

\usepackage{natbib}
\bibpunct{(}{)}{;}{a}{}{,}

\usepackage{amsmath}
\usepackage{comment}
\usepackage{amssymb}
\usepackage{xspace}
\usepackage{graphicx}

\newcommand{\Mag}{\ensuremath{{M_\lambda}}}
\newcommand{\enh}{\ensuremath{{[\alpha\text{/Fe}]}}}
\newcommand{\metal}{\ensuremath{Z}}
\newcommand{\age}{\ensuremath{\tau}}
\newcommand{\mass}{\ensuremath{m}}
\newcommand{\T}{\ensuremath{T_{\text{eff}}}}

\newcommand{\FeH}{\ensuremath{{\text{[Fe/H]}}}\xspace}
\newcommand{\MH}{\ensuremath{{\text{[M/H]}}}\xspace}
\newcommand{\mH}{\ensuremath{{\text{[m/H]}}}}

\newcommand{\jk}{{\ensuremath{(J-K_s)}}}
\newcommand{\logT}{\ensuremath{\log(\T)}}
\newcommand{\logg}{\ensuremath{\log(g)}}
\newcommand{\kms}{\ensuremath{{\rm km\,s}^{-1}}\xspace}
\newcommand{\avg}[1]{\overline{#1}}

\newcommand{\F}{\mathcal{F}}
\newcommand{\Sf}{\mathcal{F}_0}
\newcommand{\I}{\mathcal{I}}

\RequirePackage[normalem]{ulem}

\newcommand{\raveInput}{51\,829\xspace}
\newcommand{\raveApp}{22\,407\xspace}
\newcommand{\ravePre}{16\,645\xspace}
\newcommand{\raveClean}{16\,146\xspace}
\newcommand{\raveApply}{16\,146\xspace}
\newcommand{\raveApplyMS}{6\,975\xspace}

\newcommand{\raveMaga}{12\,110\xspace}
\newcommand{\raveMagaperc}{50}
\newcommand{\raveMagb}{8\,073\xspace}
\newcommand{\raveMagbperc}{45}
\newcommand{\raveMagc}{4\,037\xspace}
\newcommand{\raveMagcperc}{35}

\newcommand{\raveMagaMS}{5\,231\xspace}
\newcommand{\raveMagaMSperc}{42}
\newcommand{\raveMagbMS}{3\,488\xspace}
\newcommand{\raveMagbMSperc}{36}
\newcommand{\raveMagcMS}{1\,744\xspace}
\newcommand{\raveMagcMSperc}{31}

\newcommand{\raveVel}{5\,020\xspace}
\newcommand{\raveVelVol}{3\,249\xspace}

\begin{document}
\title{Distance determination for RAVE stars using stellar models}
\titlerunning{Distance determination for RAVE stars}

\author{
   M.A.~Breddels\inst{1},
   M.C.~Smith\inst{2,3,1},
   A.~Helmi\inst{1},
   O.~Bienaym\'e\inst{4},
   J.~Binney\inst{5},
   J.~Bland-Hawthorn\inst{6},
   C.~Boeche\inst{7},
   B.C.M.~Burnett\inst{5},
   R.~Campbell\inst{7},
   K.C.~Freeman\inst{8},
   B.~Gibson\inst{9},
   G.~Gilmore\inst{3},
   E.K.~Grebel\inst{10},
   U.~Munari\inst{11},
   J.F.~Navarro\inst{12},
   Q.A.~Parker\inst{13},
   G.M.~Seabroke\inst{14},
   A.~Siebert\inst{4},
   A.~Siviero\inst{11,7},
   M.~Steinmetz\inst{7},
   F.G.~Watson\inst{6},
   M.~Williams\inst{7},
   R.F.G.~Wyse\inst{15},
   T.~Zwitter\inst{16}
}
\authorrunning{M.A.~Breddels et al.}

\institute{
Kapteyn Astronomical Institute, University of Groningen, P.O. Box 800,
9700 AV Groningen, the Netherlands\and
Kavli Institute for Astronomy and Astrophysics, Peking University, Beijing 100871, China\and
Institute of Astronomy, University of Cambridge, Cambridge, UK\and
Universit\'e de Strasbourg, Observatoire  Astronomique,  Strasbourg,
France \and
Rudolf Peierls Centre for Theoretical Physics, Oxford, UK\and
Anglo-Australian Observatory, Sydney, Australia\and
Astrophysikalisches Institut Potsdam, Potsdam, Germany\and
RSAA, Australian National University, Canberra, Australia\and
University of Central Lancashire, Preston, UK\and
Astronomisches Rechen-Institut, Center for Astronomy of the University
of Heidelberg, Heidelberg, Germany\and
INAF, Astronomical Observatory of Padova, Asiago station, Italy\and
University of Victoria, Victoria, Canada\and
Macquarie University, Sydney, Australia\and
e2v Centre for Electronic Imaging, Planetary and Space Sciences
Research Institute, The Open University, Milton Keynes, UK\and
Johns Hopkins University, Baltimore, MD, USA\and
Faculty of Mathematics and Physics, University of Ljubljana,
Ljubljana, Slovenia
}

\abstract{} {We develop a method for deriving distances from
spectroscopic data and obtaining full 6D phase-space coordinates for the
RAVE survey's second data release.}  {We used stellar models combined with
atmospheric properties from RAVE (effective temperature, surface
gravity and metallicity) and $\jk$ photometry from archival sources to
derive absolute magnitudes. In combination with apparent magnitudes,
sky coordinates, proper motions from a variety of sources and radial
velocities from RAVE, we are able to derive the full 6D phase-space
coordinates for a large sample of RAVE stars.  This method is tested
with artificial data, Hipparcos trigonometric parallaxes and
observations of the open cluster M67.} {When we applied our method
to a set of $\raveApply$ stars, we found that 25\% ($\raveMagc$) of the stars have relative (statistical)
distance errors of $< \raveMagcperc$\%, while 50\% ($\raveMagb$) and 75\% ($\raveMaga$) have relative
(statistical) errors smaller than \raveMagbperc{}\% and \raveMagaperc{}\%, respectively.  Our
various tests show that we can reliably estimate distances for
main-sequence stars, but there is an indication of potential
systematic problems with giant stars owing to uncertainties in the
underlying stellar models. For the main-sequence star sample (defined
as those with $\logg{} > 4$), 25\% ($\raveMagcMS$) have
relative distance errors $< \raveMagcMSperc\%$, while 50\% ($\raveMagbMS$)
and 75\% ($\raveMagaMS$) have
relative errors smaller than \raveMagbMSperc{}\% and \raveMagaMSperc{}\%, respectively. Our
full dataset shows the expected decrease in the metallicity of stars
as a function of distance from the Galactic plane. The known kinematic
substructures in the $U$ and $V$ velocity components of nearby dwarf
stars are apparent in our dataset, confirming the accuracy of
our data and the reliability of our technique. We provide independent
measurements of the orientation of the $UV$ velocity ellipsoid and of
the solar motion, and they are in very good agreement with previous
work.}{The distance catalogue for
the RAVE second data release is available at {\sf
http://www.astro.rug.nl/$\sim$rave}, and will be updated in the future
to include new data releases.}

\keywords{Methods: numerical - Methods: statistical - Stars: distances - Galaxy: kinematics and dynamics - Galaxy: structure}

\maketitle

\section{Introduction}

The spatial and kinematic distributions of stars in our Galaxy contain
a wealth of information about its current properties, its history and
evolution.
This phase-space distribution is a crucial ingredient if we are to
build and test dynamical models of the Milky Way \citep[e.g.][and
references therein]{Bi2005}. More directly, the kinematics of halo
stars can be used to trace the Galaxy's accretion history
\citep{HelmiWhite1999},
as has been shown to good effect in many subsequent studies
\citep[e.g.][]{Helmi1999,Kepley2007,Smith2009}.
There is also much to learn from the phase-space structure of the
disk, where it is possible to identify substructures due to both
accretion events and dynamical resonances
\citep[e.g.][]{Dehnen2000,Famaey2005,Helmi2006} or learn about the
mixing processes that influence the chemical evolution of the disk
\citep[e.g.][]{Ro2008,Sch2008}.

To fully exploit this rich resource, we need to analyse the full
six-dimensional phase-space distribution, which clearly cannot be done
without a reliable estimate of the distances to the stars under
consideration. Therefore obtaining accurate distances and velocities
for a representative sample of stars in our Galaxy will be essential
if we are to understand both the structure of our own Galaxy and
galaxy formation in general.

The most dramatic recent development in this field was the Hipparcos
satellite mission \citep{ESA1997,2000A&A...355L..27H}, which carried
out an astrometric survey of stars down to $V \sim 12$ mag with
accuracies of up to 1 mas. This catalogue enabled the distances of
$\sim 10,000$ stars to be measured using the trigonometric
parallax technique, with parallax errors of less than 5\%
\citep{vanLeeuwen2007a,vanLeeuwen2007b}.
However, in general the resulting parallaxes only probe out to a
couple of hundred parsec and are limited to the brightest stars.

This limitation of the trigonometric parallax method led researchers
to attempt other techniques for calculating distances. One promising
avenue is the study of pulsating variable stars, such as RR Lyraes or
Cepheids, for which it is possible to accurately determine distances
using period-luminosity relations \citep[see, for example, the reviews
of][]{Ga1995,Ga1996}.  These have been used effectively to probe the
structure of our Galaxy, in particular the study of the old and
relatively metal-poor RR Lyrae stars \citep{Vi2001,Ku2008,Wa2009}.

Although pulsating variables can provide accurate tracer populations,
the numbers of such stars is clearly limited; ideally we would like to
determine distances for large numbers of stars and not just specific
populations. As a consequence there have been numerous studies
utilising photometric distance determinations, where one estimates the
absolute magnitude of a star from its colour. The efficacy of this
method can be seen from the work of \citet{Siegel2002} and
\citet{Juric2008}, who both used this technique to model the
stellar density distribution of the Galaxy. Another striking example
of the power of this technique was presented by \citet{Belokurov2006},
where halo turn-off stars were used to illuminate a host of
substructures in the Galactic halo.

The strength of photometric distances is that they can be
constructed for a wide range of stellar populations. An important
recent study was carried out by \citet{Ivezic2008}. In this work they
took high-precision multi-band optical photometry from the Sloan
Digital Sky Survey \citep[SDSS;][]{Ab2008} and constructed a
photometric distance relation for F- and G-type dwarfs, using colours
to identify main-sequence stars and estimate metallicity. Globular
clusters were used to calibrate their photometric relation,
resulting in distance estimates accurate to $\sim15$ per cent. This is
only possible due to the extremely well-calibrated SDSS photometry
and, in any case, is only applicable to F- and G-type dwarfs.
To determine distances for entire surveys (with a wide range of different
stellar classes and populations) requires complex multi-dimensional
algorithms. In this paper we develop such a technique to estimate
distances for stars using photometry in combination with stellar
atmosphere parameters derived from spectra.

One of the motivations behind our study is so that we can complement
the Radial Velocity Experiment
\citep[RAVE][]{Steinmetz2006,Zwitter2008}. This project, which started
in 2003, is currently measuring radial velocities and stellar
atmosphere parameters (temperature, metallicity and surface gravity)
for stars in the magnitude range 9~$<$~$I$~$<$~12. By the time it
reaches completion in $\sim2011$ it is hoped that RAVE will have
observed up to one million stars, providing a dataset that will be of
great importance for Galaxy structure studies. A number of
publications have already made use of this dataset
\citep[e.g.][]{Smith2007, Klement+2008,Munari2008,Siebert+2008,Veltz2008},
but to fully utilise the kinematic information we crucially need to
know the distances to the stars. Unfortunately, most of the
stars in the RAVE catalogue are too faint to have accurate
trigonometric parallaxes, hence the importance of a reliable and
well-tested photometric/spectroscopic parallax algorithm.
When distances are combined with archival proper motions and high
precision radial velocities from RAVE, this dataset will provide the
full 6D phase-space coordinates for each star.
Clearly such an algorithm for estimating distances will be a vital
tool when carrying out kinematic analyses of large samples of Galactic
stars, not just for the RAVE survey but for any similar study.

The future prospects for distance determinations are very
promising. In the next decade the Gaia satellite \citep{Perryman2001}
will observe up to $10^{9}$ stars with exquisite astrometric
precision. The mission is due to start in 2012, but a
final data release will not arrive until near the end of the decade at
the earliest.
Furthermore, as with any such magnitude limited survey, there
will be a significant proportion of stars for which their distances
are too great for accurate trigonometric parallaxes to be
determined. Therefore, although Gaia will revolutionise this
field, it will not close the chapter on distance determinations for
stars in the Milky Way and so photometric parallax techniques will
remain of crucial importance.

In this paper we present our algorithm for determining distances,
which we construct using stellar models. When we apply this method to
the RAVE dataset we are able to reproduce several known
characteristics of the kinematics of stars in the solar neighbourhood.
In \S \ref{sec:method}, we present
a general introduction. We discuss the connection between stellar
evolution theory, stellar tracks and isochrones to gain insight in
these topics before presenting our statistical methods for the
distance determination and testing the method using synthetic data. In
\S \ref{sec:rave} we apply the method to the RAVE dataset and compare
the distances to external determinations, namely stars in the open
cluster M67 and nearby stars with trigonometric parallaxes from
Hipparcos. Results obtained from the phase-space distribution are
presented in \S \ref{sec:results} to check whether the data reflect
known properties of our Galaxy. We present a discussion of the
uncertainties and limitations of the method in \S \ref{sec:discussion}
and conclude with \S \ref{sec:conc}.

%%%%%%%%%%%%%%%%%%%%%%%%%%%%%%%%%%%%%%%%%%%%%%%%%%%%%%%%%%%%%%%%%%%%%%%
%%%%%%%%%%%%%%%%%%%%%%%%%%%%%%%%%%%%%%%%%%%%%%%%%%%%%%%%%%%%%%%%%%%%%%%
%%%%%%%%%%%%%%%%%%%%%%%%%%%%%%%%%%%%%%%%%%%%%%%%%%%%%%%%%%%%%%%%%%%%%%%
%%%%%%%%%%%%%%%%%%%%%%%%%%%%%%%%%%%%%%%%%%%%%%%%%%%%%%%%%%%%%%%%%%%%%%%
%%%%%%%%%%%%%%%%%%%%%%%%%%%%%%%%%%%%%%%%%%%%%%%%%%%%%%%%%%%%%%%%%%%%%%%
%%%%%%%%%%%%%%%%%%%%%%%%%%%%%%%%%%%%%%%%%%%%%%%%%%%%%%%%%%%%%%%%%%%%%%%
%%%%%%%%%%%%%%%%%%%%%%%%%%%%%%%%%%%%%%%%%%%%%%%%%%%%%%%%%%%%%%%%%%%%%%%
%%%%%%%%%%%%%%%%%%%%%%%%%%%%%%%%%%%%%%%%%%%%%%%%%%%%%%%%%%%%%%%%%%%%%%%
%%%%%%%%%%%%%%%%%%%%%%%%%%%%%%%%%%%%%%%%%%%%%%%%%%%%%%%%%%%%%%%%%%%%%%%
%%%%%%%%%%%%%%%%%%%%%%%%%%%%%%%%%%%%%%%%%%%%%%%%%%%%%%%%%%%%%%%%%%%%%%%
%%%%%%%%%%%%%%%%%%%%%%%%%%%%%%%%%%%%%%%%%%%%%%%%%%%%%%%%%%%%%%%%%%%%%%%
%%%%%%%%%%%%%%%%%%%%%%%%%%%%%%%%%%%%%%%%%%%%%%%%%%%%%%%%%%%%%%%%%%%%%%%

\section{Method for distance determination}
\label{sec:method}
\subsection{Stellar models and observables}
\label{sec:method:se}

Stellar models are commonly used to estimate distances, for instance in main-sequence
fitting. Such methods work for collections of stars, but models
can also be used to infer properties of individual stars, such as ages
\citep{PontEyer2004,Jorgensen2005,daSilva2006}. In our analysis we utilise
this approach, combining stellar parameters (temperature, metallicity
and surface gravity) with photometry to estimate a star's absolute
magnitude.

The evolution of a star is fully determined by its mass and initial
chemical composition \citep[e.g.][]{2005essp.book.....S}.
Stellar tracks and isochrones can be
seen (in a mathematical sense) as a function ($\F$) of
alpha-enhancement ([$\alpha$/Fe]), metallicity ($\metal$), mass
($\mass$) and age ($\age$) that maps onto the observables: absolute
magnitude ($\Mag$), surface gravity ($\logg$), effective temperature
($\T$), and colours, i.e.
\begin{equation}
	\label{eq:isochrone1}
	\begin{array}{ll}
	& \F(\enh, \metal, \age, \mass) \mapsto (\Mag, \logg, \T, \text{colours,\ldots} ). \\
	\end{array}
\end{equation}
In particular, an isochrone is the function $\I(\mass)$ of mass, which is obtained from $\F$ by keeping all other variables constant.

Assuming solar $\alpha$-abundance, \enh = 0, we define the function $\Sf(\metal, \age, \mass)$, which is $\F$ with $\enh$ fixed at 0,
\begin{equation}
	\label{eq:S}
	\begin{array}{lll}
	\Sf(\metal, \age,\mass) & = & \F(\metal, \age,\mass)|_{\enh=0} \\
	& \mapsto & (\Mag, \logg, \T, {\rm colours}... ).\\
	\end{array}
\end{equation}
Therefore the isochrones or stellar tracks from a given model can be
seen as samples from the theoretical stars defined by $\Sf(\metal,
\age, \mass)$. Throughout this paper we assume solar-scaled
metallicities, which means that $\enh=0$ and $\MH=\FeH$, where $\MH$
is defined as $\log(Z/Z_\odot)$.

%%%%%%%%%%%%%%%%%%%%%%%%%%%%%%%%%%%%%%%%%%%

For our study we use the $Y^2$ (Yonsei-Yale) models
\citep{Demarque2004}. These models can be downloaded from the $Y^2$
website\footnote{{http://www-astro.physics.ox.ac.uk/\~{}yi/yyiso.html}},
where also an interpolation routine is available, called YYmix2.
It should be noted that these models ignore any element diffusion
that may take place in the stellar atmosphere \citep[see, for example,][]{Toma2008}.

\begin{figure}
	\centerline{\includegraphics[scale=0.90]{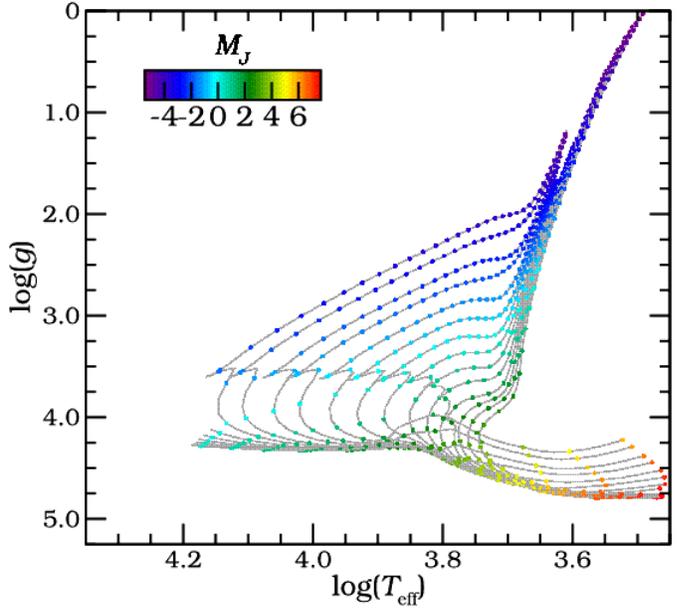}}
	\caption{
		$\logg$ versus $\log(\T)$ plot for isochrones from 0.01-15 Gyr spaced logarithmically, for \MH = 0 and \enh\ = 0. Colour indicates the absolute magnitude in the $J$ band.
		\label{fig:colorisochrones}
	}
\end{figure}

\begin{figure}
	\centerline{\includegraphics[keepaspectratio=true,scale=0.9]{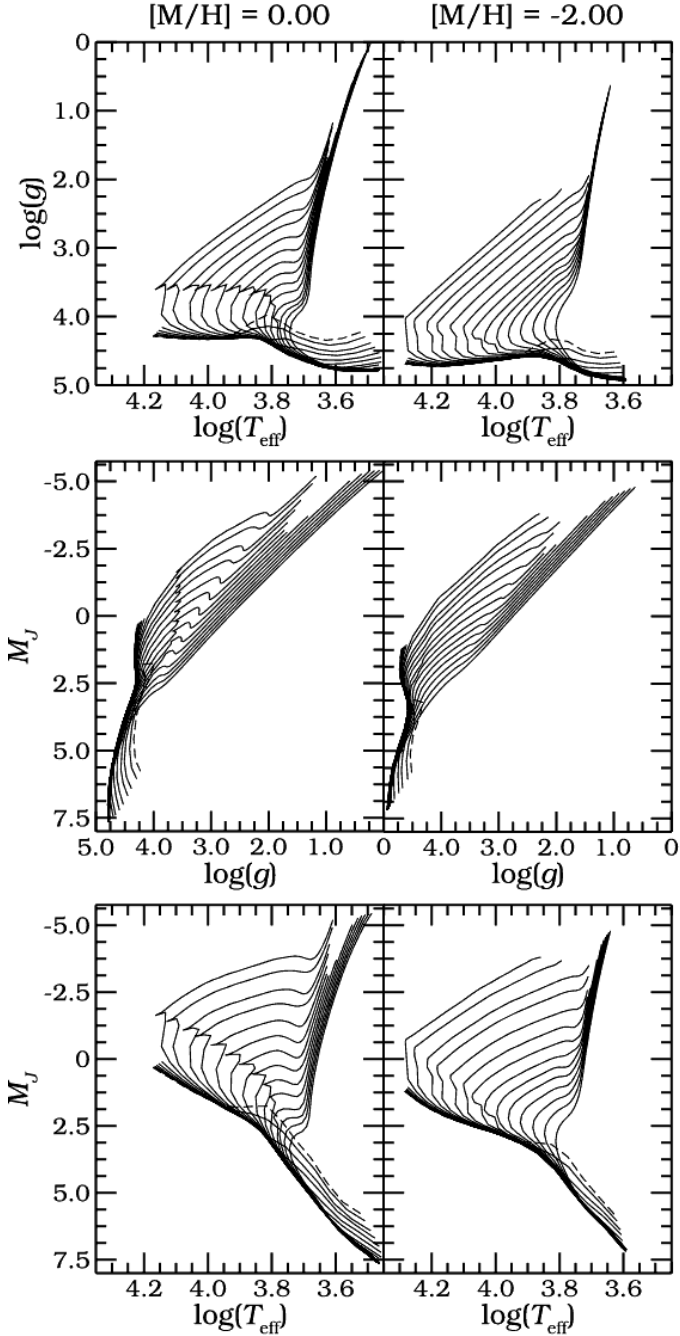}}
	\caption{
		Isochrones for \enh{} = 0, \MH{} = 0 (left column)
                and \MH{} = $-2$ (right column), ages ranging from
                0.01-15 Gyr spaced logarithmically.  The dashed line
                indicates the youngest (0.01 Gyr) isochrone. {\bf Top
                  row}: Similar to Fig. \ref{fig:colorisochrones},
                shown for completeness. {\bf Middle row}: $M_J$ is
                best restricted by $\logg$ for RGB stars. {\bf Bottom
                  row}: $M_J$ is best restricted by $\T$ for
                main-sequence stars.
	}\label{fig:isochrones}
\end{figure}

A sample of theoretical `model stars' from these
$Y^2$ models are shown in Fig. \ref{fig:colorisochrones}. Each model
star is represented as a dot and the connecting lines correspond to
the isochrones of different ages. In Fig. \ref{fig:isochrones} we
show the same isochrones as Fig. \ref{fig:colorisochrones},
illustrating the relation between $M_J$ and $\T$, and between $M_J$
and $\logg$ separately. Clearly, for a given $\T$, $\logg$ and \MH
it is not possible to infer a unique $M_J$ (i.e. the function $\Sf$ is not injective). This can be seen most
clearly in Fig. \ref{fig:colorisochrones}, where around \logT{} = 3.8,
\logg{} = 4 the isochrones overlap. However, this is also evident in
other regions; for example in the top panels of
Fig. \ref{fig:isochrones} the isochrones are systematically
shifted as metallicity goes from 0 to $-2$. Because we are unable to
determine a unique $M_J$ for a given star we are
forced to adopt a statistical approach, i.e. obtaining a probability
distribution for $M_J$.

From Fig. \ref{fig:isochrones} we can see how errors in the observables $\logg$ and $\T$
affect the uncertainty in the absolute magnitude ($M_J$ in this
example). The middle row in Fig. \ref{fig:isochrones} shows that the
value of $M_J$ is better defined by $\logg$ for red giant branch (RGB)
stars than for main-sequence stars, independently of their
metallicity. On the other hand, the bottom row of
Fig. \ref{fig:isochrones} shows that $\T$ essentially
determines $M_J$ for main-sequence stars, again independently of
metallicity. We therefore expect that a small error in $\logg$ will
give better absolute magnitude estimates for RGB stars, while a small
error in $\T$ will have a similar effect on main-sequence stars. We
also expect this not to be strongly dependent on metallicity.

\subsection{Description of the method}
\label{sec:method:method}

We now outline the method that we use to estimate the probability
distribution function (PDF) for the absolute magnitude (or, equivalently,
the distance). Previous studies have employed similar techniques to
determine properties of stars using stellar models. A selection of
such work can be found in the following references:
\citet{PontEyer2004,Jorgensen2005,daSilva2006}.

Our method requires
a set of model stars. As was discussed in \S \ref{sec:method:se}, we
have chosen to use the $Y^2$ models \citep{Demarque2004}. We generate
our set of isochrones using the YYmix2 interpolation code. The set
consists of 600 isochrones, with 40 different ages, spaced
logarithmically between 0.01 and 15.0 Gyr, and 15 different
metallicities with 0.25 dex separation (corresponding to 1 sigma in
\MH{} for the RAVE data; see Section \ref{sec:rave:data}) between
\MH{} = $-2.5$ and \MH{} = $1.0$. The separation between the points of
the isochrones has been visually inspected and is, in general, smaller
than the errors in \T{} and \logg{}. These isochrones do not track the
evolution beyond the RGB tip. We only use the isochrones with \enh{} =
0 because our observational data do not allow an accurate measurement
of \enh{} and for most of our stars we expect $\enh\approx0$.
Later, in \S \ref{sec:method:test}, we show that assuming $\enh = 0$
for stars having $\enh > 0$ does not introduce any noticeable bias in
our results.

Let us suppose we have measured the following parameters for a sample
of stars: $\T$, $\logg$, \MH{} and $\jk$. Each of these quantities
will have associated uncertainties due to measurement errors
($\sigma_{\T}$, $\sigma_{\logg}$, $\sigma_{\MH}$ and
$\sigma_{\jk}$), which we assume are Gaussian.
For each observed star we first need to obtain the closest
matching model star, which we do by minimising the usual $\chi^2$ statistic,
\begin{equation}
\chi^2_{\rm model} = \sum_{i=1}^{n} \frac{(A_i-A_{i,{\rm model}})^2}{\sigma_{A_i}^2}
\label{eq:chisq_model},
\end{equation}
where $A_i$ corresponds to our observable parameters
(i.e. $n=4$ in this case) and $A_{i,{\rm model}}$ the corresponding
parameters of the model star, as given by the set of isochrones.
By minimising Eq. (\ref{eq:chisq_model}), we obtain the parameters for
the most-likely model star, denoted $\avg{A_1},...,\avg{A_n}$.

Having identified the most probable model, we generate 5000
realisations of the observations that could be made of this model star
by sampling Gaussian distributions in each observable that are centred
on the model values, with the dispersion in each observable equal to
the errors in that quantity.\footnote{Note that since $J$ comes into
the method twice (once for $(J-K_s)$ and once in the distance modulus),
we draw $J$ and $K_s$ separately to ensure that the correlations are
treated correctly.}  By drawing our realisations about $\avg{A_i}$ we
are making the assumption that the observables are just a particular
realisation of the model \citep[e.g. chapter 15.6 of][]{Press1992}.
Then for each such realisation we again find the most probable star by
minimising $\chi^2_{\rm model}$ in Eq. (\ref{eq:chisq_model}). The
final PDF is the frequency distribution of the intrinsic properties of
the model stars that have been located in this way. One may argue
that the first step of finding the closest model star is not formally
correct since it does not have a corresponding Bayesian
equivalent. However, we have found no apparent differences in the
results in tests where we exclude this step in the procedure.

We use the PDF obtained from the Monte Carlo realisations to determine
the distance.  Due to the non-linearity of the isochrones, as can be
seen in Fig. \ref{fig:isochrones}, we expect the PDFs to be
asymmetric. In such cases the mode and the mean of the PDF are not the
same. Since the mean is a linear function,\footnote{The mean of a set
of means is equal to the mean of the combined PDF.} we choose to
calculate the mean and standard deviations of $M_J$ (and distance $d$) from the
Monte Carlo realisations. This gives us our final determination for
the distance to each star and its associated error. We also
compared the method using the median of the distribution of absolute
magnitudes instead of the mean, and found no significant differences.

We have not made use of any priors in this analysis. We could have
invoked a prior based on, for example, the luminosity function or mass
function of stars in the solar neighbourhood.
However, since the luminosity function of our sample is not an
unbiased selection from the true luminosity function in the RAVE
magnitude range \citep{Zwitter2008}, this makes the task of
quantifying our prior very difficult. We therefore choose to adopt a
flat prior in order to avoid any potential biases from incorrect
assumptions. However, it is hoped that by the end of the RAVE survey it will have produced a magnitude limited catalogue, at which point it may become possible to invoke a prior based on the luminosity function.

\subsection{Testing the method}
\label{sec:method:test}
To test the method, we take a sample of 1075 model stars. This
set is large enough for testing purposes, allowing us to determine
which kind of stars the method works best for.
The sample of 1075 model stars
are taken from a coarsely generated grid of isochrone models with metallicity $\MH = 0$.
We convolve \MH{}, $\T$,
$\logg$ and the colours with Gaussians with dispersions
comparable to the error in the RAVE survey in order to mimic our
measurements ($\sigma_{\T}$ = 300 K, $\sigma_{\logg}$ = 0.3 dex,
$\sigma_{\MH}$ = 0.25 dex, $\sigma_J \approx \sigma_{K_s} \approx
$ 0.02 mag; see \S \ref{sec:rave:data}).

The reason for choosing a fixed metallicity is twofold. In \S
\ref{sec:method:se} we have seen that different metallicities should
give similar results in terms of the precision with which the absolute
magnitude can be derived. Secondly, it also means that the results
only have to be compared to one set of isochrones, making it easier to
interpret. Note that although one metallicity is used to generate the
sample, after error convolution, isochrones for all metallicities are
used for the fitting method.

\begin{figure}[!htp]
\centerline{\includegraphics[keepaspectratio=true,scale=0.6]{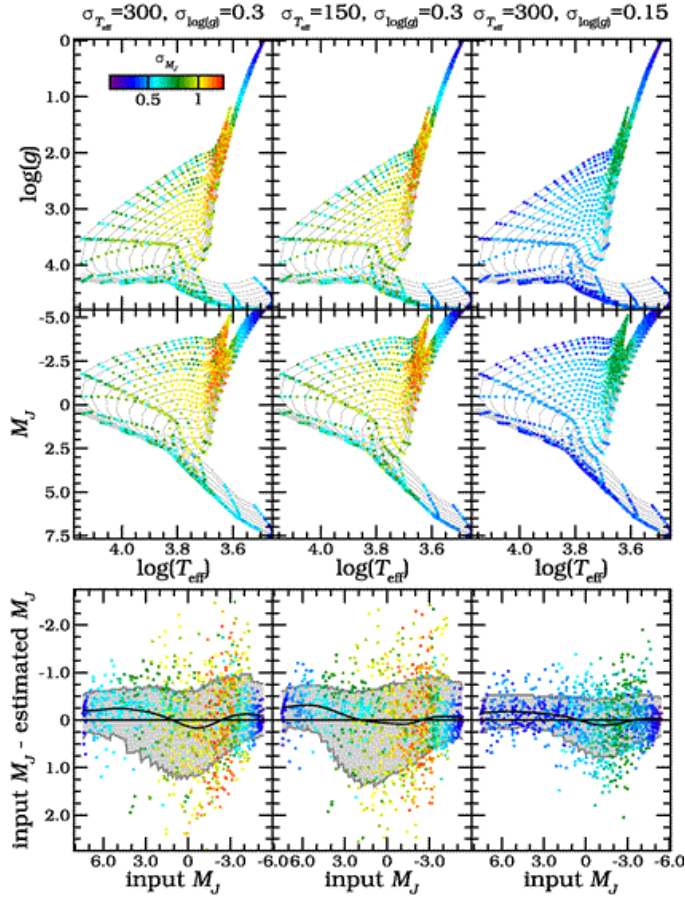}}
\caption{
	Effect of the uncertainties in $\logg$ and $\T$ on the estimated
        absolute magnitude $M_J$. The main-sequence and RGB stars
        perform best. Reducing the errors in $\logg$ has the largest
        effect. {\bf Left column}: Errors similar to the RAVE dataset,
        $\sigma_{\T}$ 300 K and $\sigma_{\logg}$ = 0.3. {\bf Middle
          column}: Reducing the errors in effective temperature,
        $\sigma_{\T}$ = 150 K. {\bf Right column}: Reducing the errors in
        surface gravity, $\sigma_{\logg}$ = 0.15. {\bf Top row}: The
        sample of 1075 stars, with colours indicating errors,
        clipped to a value of $\sigma_{M_J}$ = 1.25. {\bf Middle row}:
        CMD with colours indicating the same errors as the top
        row. {\bf Bottom row}: Difference between input (i.e. model)
        and estimated absolute magnitude versus input absolute
        magnitude. We include a running mean and dispersion. The
        colours correspond to the same scale as in the top row. The
        spread in this distribution grows as the estimated uncertainty
        in $M_J$ grows (as indicated by the colour change).
 }\label{fig:mjvsmj}
\end{figure}

We run the method described in the previous section on this set of
1075 stars and analyse the results in the left column of
Fig. \ref{fig:mjvsmj}. The colours indicate the estimated errors on
$M_J$ obtained from our algorithm and are clipped to a value
$\sigma_{M_J}$ = 1.25. The middle row shows the results on a
colour-magnitude diagram (CMD). Stars
on the main sequence and on the RGB appear to have the smallest errors
as expected (see \S \ref{sec:method:se}). In the bottom row, the
difference between the input (i.e. model) and
estimated magnitude is plotted against the input
magnitude of the model star from which the estimate was derived,
showing the deviation from the input absolute magnitude grows with
$\sigma_{M_J}$, as expected. The method appears to give reasonable
results, showing no serious systematic biases. The left column of
Fig. \ref{fig:mjvsmj} shows that for the main sequence and
RGB stars in the RAVE data set we expect a relative distance error of the
order of 25\% (blue colours), and for the other stars around 50-60\%
(green to red colours).

We run this procedure again, now testing the effect of reducing the
error in $\T$. If we decrease the error in $\T$ to 150 K, we obtain the results shown in the middle column of
Fig. \ref{fig:mjvsmj}. The errors in $M_J$ do not seem to have changed
much, except for a very slight improvement for the main-sequence
stars. If, on the other hand, we decrease the error in $\logg$ to 0.15
dex while keeping the $\T$ error at 300 K, we obtain the results shown
in the right column in Fig. \ref{fig:mjvsmj}. This shows that the
accuracy and precision with which we can determine $M_J$ has increased
significantly. Therefore, reducing the uncertainty in $\logg$ is much
more effective than a similar reduction in $\T$ and will result in
significant improvements in the estimate of the absolute magnitude. In
future, high precision photometry from surveys such as Skymapper's
Southern Sky Survey \citep{Ke2007} may aid the ability of RAVE to
constrain the stellar parameters.

We carry out an additional test to quantify whether our decision to
only fit to $\enh=0$ models will bias our results. To do this we
generated three similar catalogues of model stars, but with
$\enh{}=0,0.2,0.4$ dex. We then repeat the above procedure (as usual
fitting to models with \enh{} fixed at 0) and analyse the resulting
distances. Reassuringly we find that there is no difference between
the accuracy of the three catalogues, justifying our decision to carry
out the model fitting using only $\enh=0$ models.

%%%%%%%%%%%%%%%%%%%%%%%%%%%%%%%%%%%%%%%%%%%%%%%%%%%%%%%%%%%%%%%%%%%%%%%
%%%%%%%%%%%%%%%%%%%%%%%%%%%%%%%%%%%%%%%%%%%%%%%%%%%%%%%%%%%%%%%%%%%%%%%
%%%%%%%%%%%%%%%%%%%%%%%%%%%%%%%%%%%%%%%%%%%%%%%%%%%%%%%%%%%%%%%%%%%%%%%
%%%%%%%%%%%%%%%%%%%%%%%%%%%%%%%%%%%%%%%%%%%%%%%%%%%%%%%%%%%%%%%%%%%%%%%
%%%%%%%%%%%%%%%%%%%%%%%%%%%%%%%%%%%%%%%%%%%%%%%%%%%%%%%%%%%%%%%%%%%%%%%
%%%%%%%%%%%%%%%%%%%%%%%%%%%%%%%%%%%%%%%%%%%%%%%%%%%%%%%%%%%%%%%%%%%%%%%
%%%%%%%%%%%%%%%%%%%%%%%%%%%%%%%%%%%%%%%%%%%%%%%%%%%%%%%%%%%%%%%%%%%%%%%
%%%%%%%%%%%%%%%%%%%%%%%%%%%%%%%%%%%%%%%%%%%%%%%%%%%%%%%%%%%%%%%%%%%%%%%
%%%%%%%%%%%%%%%%%%%%%%%%%%%%%%%%%%%%%%%%%%%%%%%%%%%%%%%%%%%%%%%%%%%%%%%
%%%%%%%%%%%%%%%%%%%%%%%%%%%%%%%%%%%%%%%%%%%%%%%%%%%%%%%%%%%%%%%%%%%%%%%
%%%%%%%%%%%%%%%%%%%%%%%%%%%%%%%%%%%%%%%%%%%%%%%%%%%%%%%%%%%%%%%%%%%%%%%
%%%%%%%%%%%%%%%%%%%%%%%%%%%%%%%%%%%%%%%%%%%%%%%%%%%%%%%%%%%%%%%%%%%%%%%

\section{Application to RAVE data}
\label{sec:rave}

\subsection{Data}
\label{sec:rave:data}

The Radial Velocity Experiment (RAVE) is an ongoing project
measuring radial velocities and stellar atmosphere parameters
(temperature, metallicity, surface gravity and rotational velocity) of
up to one million stars in the Southern hemisphere. Spectra are taken
using the 6dF spectrograph on the 1.2m UK Schmidt Telescope of the
Anglo-Australian Observatory, with a resolution of R = 7\,500, in the
$8\,500-8\,750$ \AA\xspace window.
The input catalogue has been constructed from the Tycho-2 and
SuperCOSMOS catalogues in
the magnitude range $9 < I < 12$.
To date RAVE has obtained spectra of over 250\,000 stars, 50\,000 of which
have been presented in the most recent data release \citep{Zwitter2008}.

This second RAVE data release provides metallicity
($\MH{}$), $\logg$ and $\T$ from the spectra, and has been
cross-matched with 2MASS to provide $J$ and $K_s$ band
magnitudes. The $(JK)_\text{ESO}$ colours used for the $Y^2$ isochrones match the 2MASS $(JK_s)_\text{2MASS}$ colours very well, so no colour transformation is required \citep{Carpenter2001}.

We choose to use the $J$ and $K_s$ bands because they are in the
infrared (IR) and are therefore less affected by dust than visual
bands. To see whether extinction will be significant for our sample we
carry out a simple test using the dust maps of \citet{Sch1998}. If we
model the dust as an exponential sheet with scale-height 130pc
\citep{Dri2001}, we find that given the RAVE field-of-view, a typical
RAVE dwarf located 250pc away would suffer $\sim$0.03 mag of
extinction in the $J$-band. This corresponds to a distance error of
$\sim1\%$, which is negligible compared to the overall uncertainty
inherent in our method. Reddening is similarly unimportant, with the
same typical RAVE star suffering $\sim$0.02 mag reddening in
$\jk$. Even if we only consider fields-of-view with $\left| b \right |
< 40^\circ$ then we find that the extinction for a star at a distance
of 250pc is only 0.04 mag (with corresponding distance error of
$\sim2\%$). Note that for future RAVE data releases it may be possible
to use information from the spectra to include extinction corrections
for some individual stars \citep{Munari2008}.

The observed parameter values used for the model fitting
routine are the weighted average of the available values, where
the weight is the reciprocal of the measurement error:
\begin{equation}
X_{\rm weighted} = \frac{\sum_j w_j X_j}{\sum_j w_j}, \label{eq:weighted_avg}
\end{equation}
where $X_j$ are the measured values and $w_j = 1/\sigma_j^2$ the corresponding weight.
The error in the average is calculated as:
\begin{equation}
\sigma_{\rm weighted}^2 = \frac{1}{\sum_j 1/\sigma_j^2}. \label{eq:weighted_error}
\end{equation}

For the RAVE data, $\T$ is determined only from the spectra, i.e. not
photometrically, which means that $\T$ and $\jk$ are
uncorrelated in the sense that they are independently
 observed. Therefore we can use both $\T$ and $\jk$ in
Eq. \ref{eq:chisq_model} to obtain our distance estimate. The error in
$\jk$ is small compared to other colours, which means that adding
further colours will result in only a negligible improvement on the
uncertainty of the absolute magnitude. For this reason we only use
this one colour.

The current RAVE data release \citep{Zwitter2008} does not include
individual errors for each star's derived parameters and so
for the errors in $\MH{}$, $\T$ and $\logg$ we take 0.25 dex, 300 K
and 0.3 dex respectively. The errors in $\MH{}$ and $\T$ are
reasonable averages for different types of stars of low temperature,
as can be seen from Fig.\ 19 in \citet{Zwitter2008}. Even though our
$\logg$ error estimate is slightly smaller compared to this figure,
our results do not show evidence of an underestimation in the distance
errors (\S \ref{sec:app:testing:hipp}).
In fact, repeated observations of certain stars in the RAVE catalogue
indicate that these errors may be conservative \citep{Steinmetz2008}.
The RAVE DR2
dataset has two metal abundances, one uncalibrated, determined from
the spectra alone ($\mH{}$), and a calibrated value ($\MH{}$). The
latter is calibrated using a subset of stars with accurate metallicity
estimates and it is this value which we use in the fitting method.
As above we assume solar-scaled metallicities, which means that
$\enh=0$ and $\MH=\FeH$.

\subsection{Determining distances to RAVE stars}
\label{sec:method:rave}

We now use the data set described above to derive absolute magnitudes
using our model fitting method (see \S \ref{sec:method:method}).

The RAVE second data release \citep{Zwitter2008} contains \raveInput
observations, of which \raveApp have astrophysical parameters. We
first clean up the dataset by requiring that the stars have all
parameters required by the fitting method ($\MH{}$, $\logg$, $\T$,
$J$, $K_s$), a signal to noise ratio S2N $>$ 20, no 2MASS photometric
quality flags raised (i.e. we require `AAA') and the spectrum quality
flag to be empty to be sure we have no obvious binaries or cosmic ray
problems. Although this latter flag will eliminate clear
spectroscopic binaries (132 individual stars, 0.2\%), our sample must
suffer from binary contamination given the estimated $37\%$ binary
fraction for F and G stars in the Copenhagen-Geneva survey
\citep{Holmberg2009} or the much lower estimates $6$-$14$\% of \citet{Famaey2005}. In future the use of repeated observations
for the RAVE sample will give a better understanding of the effect of
binaries on, for instance, the T$_{\text{eff}}$ and \logg{} estimates
(Matijevic et al. 2009, in prep.).

Although most of the RAVE survey stars in this data release are
located at high latitude (with $|b|>25^\circ$), there are a limited
number of calibration fields with $|b|<10^\circ$. We remove these
low-latitude fields from our analysis since they could suffer from
significant extinction which will bias our distance estimates.

For some stars multiple observations are available, these are grouped
by their ID, and a weighted average (Eq. \ref{eq:weighted_avg}) and
corresponding error (Eq. \ref{eq:weighted_error}) for all radial
velocities are calculated. The astrophysical parameters ($\MH{}$,
$\logg$ and $\T$) have nominal errors as described in \S
\ref{sec:rave:data}. For these parameters an unweighted average is
calculated but the error in the average is kept equal to the nominal
error. The total number of independent sources matching these
constraints is \ravePre.

Once we have our clean sample of stars we first find the best model star as
described in \S \ref{sec:method:se}. If it has a $\chi^2_{\rm model} \ge
6$ (Eq. \ref{eq:chisq_model}) it is not considered
further. This last step gets rid of the $\sim 3\%$ of stars that are
not well fit by any model.

Our final sample has \raveClean sources which are used
for the model fitting method to obtain an estimate of the distance and
associated uncertainty for each star.
\begin{figure}
	\centerline{\includegraphics[keepaspectratio=true,scale=0.9]{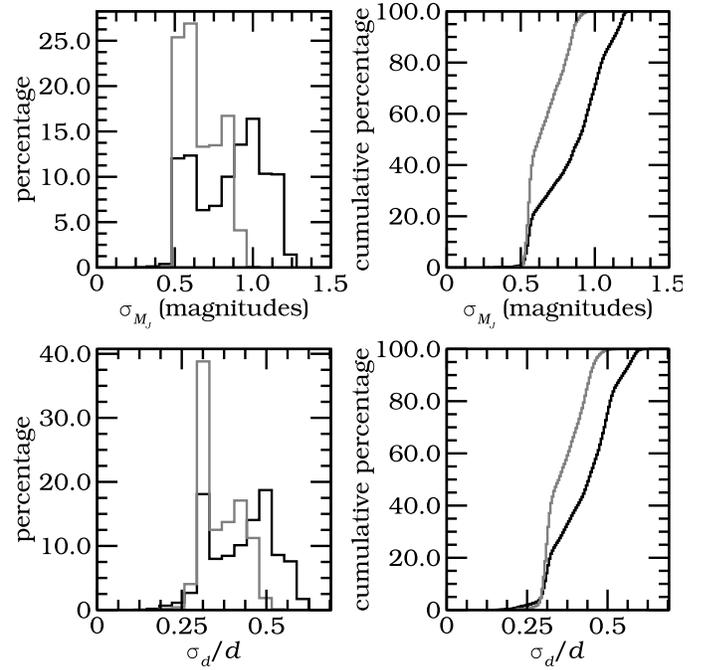}}
\caption{Error distribution (left) and cumulative plot (right) for
  $M_J$ (top) and distance (bottom). These distributions are for the
  clean sample of \raveClean stars (see \S\ref{sec:method:rave}).
  The black line includes all the stars, while the grey line shows the
  distribution for main-sequence stars (defined here as those with
  $\logg{} > 4$).}
\label{fig:errors_histograms}
\end{figure}

\begin{figure}
	\centerline{\includegraphics[keepaspectratio=true,scale=0.675]{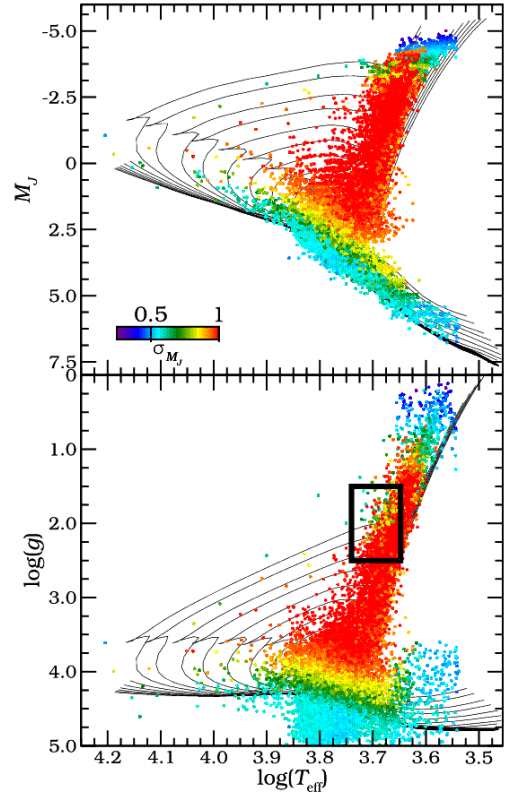}}
	%\centerline{\includegraphics[keepaspectratio=true,scale=0.675]{fig/1247105.eps}}
	\caption{Results after applying the model fitting method
	to the RAVE data. Colours indicate the magnitude of the error
	in $M_J$. Only stars with $\sigma_{M_J} < 1.0$ are
        plotted. Isochrones for \MH{} = 0 are plotted for
        comparison.
	{\bf Top}: CMD of RAVE dataset showing that the stars on the
	main sequence and RGB stars have the smallest errors.
	{\bf Bottom}: $\logT$ versus $\logg$, colour coding as in the
	top panel. The black rectangle approximately highlights the area in which red clump (RC) stars are expected to be found. The assumed error in $\logg$ is 0.3 dex and in $\T
        $ is 300 K. Note that although the RGB stars in this panel do
        not match the \MH{} = 0 isochrones, they are more consistent with the
        isochrones corresponding to their measured metallicities.}\label{fig:rave_data_errors}
\end{figure}

The distribution of uncertainties in the absolute magnitude and the
distance for this clean sample of \raveClean stars can be found in
Fig. \ref{fig:errors_histograms} (black line). The $x$-axes are scaled such that the uncertainties can be compared using
$\sigma_d/d \approx \sigma_{M_J} \ln(10)/5 = 0.46
\sigma_{M_J}$. The differences between the two histograms show
that the error in the apparent $J$ magnitude does contribute to the relative distance error. In Fig. \ref{fig:rave_data_errors} we show how the
uncertainties behave for the different types of stars.
The distribution of uncertainties for the sample is as follows:
25\% ($\raveMagc$) of the stars have relative (statistical)
distance errors of $< \raveMagcperc$\%, while 50\% ($\raveMagb$) and 75\% ($\raveMaga$) have relative
(statistical) errors smaller than \raveMagbperc{}\% and \raveMagaperc{}\% respectively. For main-sequence stars (which we define here as those with $\logg{} > 4$, the grey line in Fig. \ref{fig:errors_histograms}) the distribution of uncertainties is:
25\% ($\raveMagcMS$) have
relative distance errors $< \raveMagcMSperc\%$, while 50\% ($\raveMagbMS$)
and 75\% ($\raveMagaMS$) have
relative errors smaller than \raveMagbMSperc{}\% and \raveMagaMSperc{}\% respectively.

The $Y^2$ isochrones do not model the later evolutionary stages of stars, such as the horizontal branches and the asymptotic giant branch. The red clump (RC), which is the horizontal branch for Population I stars, is a well populated region in the CMD due to the relatively long lifetime of this phase ($\sim 0.1$ Gyr) \citep{Girardi1998}. Therefore we expect the RAVE sample to include a non negligible fraction of RC stars. Using the selection criteria of \citet{Veltz2008} and \citet{Siebert+2008}, namely $0.5 < \jk < 0.7$ and $ 1.5 < \logg < 2.5$ we find about $\sim 10$\% of the RAVE sample could be on the RC. This region is highlighted in Fig \ref{fig:rave_data_errors} with a black rectangle. The distance to many of these stars can be determined using the almost constant absolute magnitude of the RC \citep[e.g.][]{Veltz2008,Siebert+2008}. However, since there may be better ways to isolate the RC region, we choose to determine the distances for all these stars using our method. Therefore, in the rest of this paper we make no distinction between RC and RGB stars. Nonetheless, we recommend users to discard what they believe may be RC stars, and possibly to determine their distances using the absolute magnitude of the RC.

\subsection{Testing of RAVE distances}
\label{sec:app:testing}
In order to verify the accuracy of our distance estimates, we perform
two additional checks using external data and observations of the open
cluster M67.

\subsubsection{Hipparcos}
\label{sec:app:testing:hipp}

The best way to assess our distance estimates is through independent
measurements.
For calibration purposes a number of RAVE targets were chosen to be
stars previously observed by the Hipparcos mission, which means that
for these stars we will have an independent distance determination
from the trigonometric parallax. These stars are at the brighter end
of the RAVE magnitude range and are mostly dwarfs.

We take the reduction of the Hipparcos data as presented by
\citet{vanLeeuwen2007a,vanLeeuwen2007b} and cross-match these with our
RAVE stars.
In order to maximise the number of RAVE stars we use a preliminary
dataset larger than the public release described in \S
\ref{sec:rave:data}; this dataset contains $\sim250\,000$ stars, but has
not undergone the rigourous verification and cleaning of the public
data release. This cross-matching provides 624 stars for which the
Hipparcos parallax errors are less than 20\% and our distance errors
are less than 50\%. Note
that when dealing with uncertain trigonometric parallaxes it is well
known that the corresponding distance determinations are
systematically underestimated \citep{Lutz1973}. We correct for this using
the prescription described in section 3.6.2 of \citet{BM}, in
particular equation (3.51).\footnote{A mistake is present in equation
(3.51) of \citet{BM}. The correct expression can be derived from the preceding equation, which gives $\varpi/\sigma_\varpi = \left(
  \varpi'/\sigma_\varpi + \sqrt{(\varpi'/\sigma_\varpi)^2 + 4(5\beta-4)} \right)/2$, where $\varpi$ and $\varpi'$ are the true and measured parallax respectively and $\beta$ the slope of the luminosity function power law (the prior).}

\begin{figure}
	\centerline{\includegraphics[keepaspectratio=true,scale=0.8]{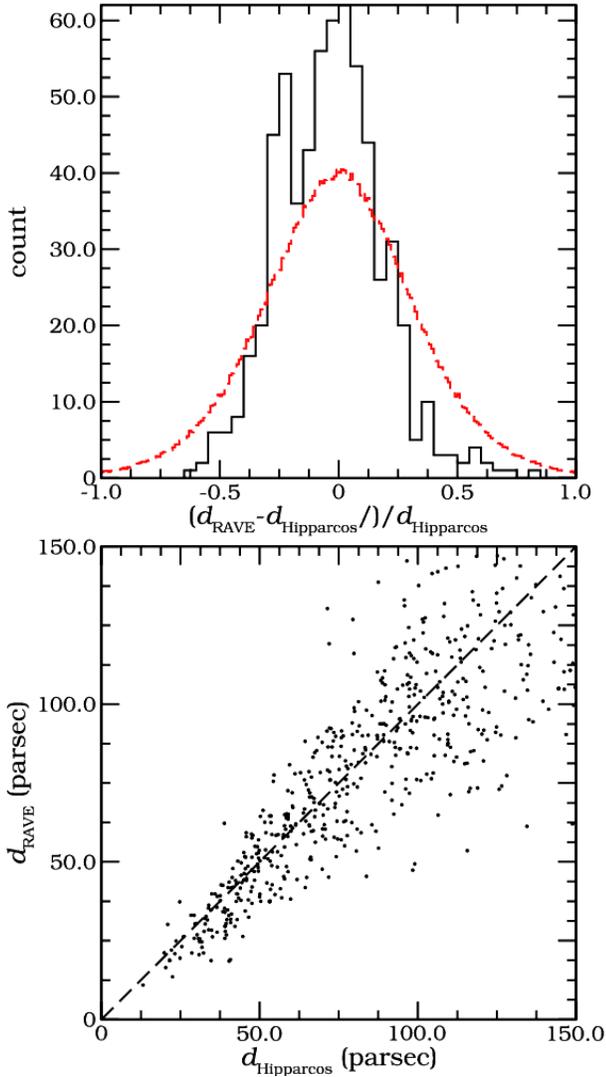}}
	\caption{{\bf Bottom:} Distance from our method versus
          Hipparcos distance, the dashed line corresponds to equal
          distances. {\bf Top:} Histogram of relative distance
          differences between our distance and that of
          Hipparcos. The dashed line shows the expected distribution
          given the quoted errors from our method and Hipparcos. Note
          that the observed distribution is narrower, indicating that
          our errors are probably overestimated for these stars (see
          \S\ref{sec:app:testing:hipp}).}\label{fig:hip_compare}
\end{figure}

\begin{figure*}
	\centerline{\includegraphics[keepaspectratio=true,scale=0.9]{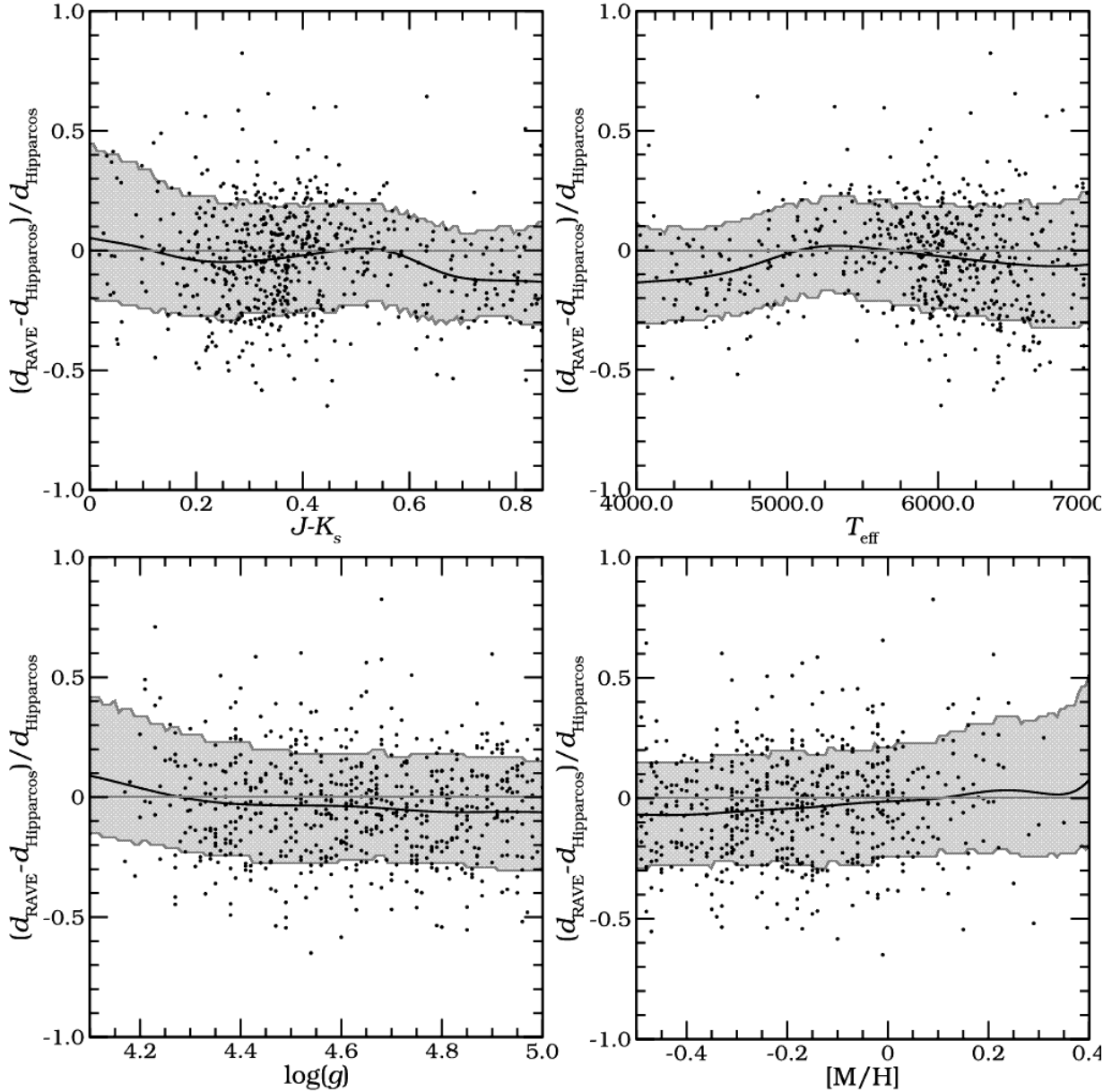}}
	\caption{Relative offset in distance from our method vs the
          trigonometric parallax determination from Hipparcos, as function of
          \logg{}, \MH{}, \jk{} and \T{}. We include a running mean
          and dispersion.}
        \label{fig:hip_systematics}
\end{figure*}

In the bottom panel of Fig. \ref{fig:hip_compare} we show a plot of our distance estimate
($d_\text{RAVE}$) vs the Hipparcos distance ($d_\text{Hipparcos}$). Clearly there
is some scatter in this distribution, but in the top panel we
quantify this by showing the distribution of $(d_\text{RAVE} -
d_\text{Hipparcos})/d_\text{Hipparcos}$. The curve shows the expected
distribution given our error on $d_\text{RAVE}$ and approximating the
error on $d_\text{Hipparcos}$ from the error on the
parallax (the true error on $d_\text{Hipparcos}$ is non-trivial to
calculate owing to the aforementioned Lutz-Kelker bias). It can be
seen that the predicted distribution is broader than the observed one;
if we assume our estimate of the Hipparcos errors are reasonable, this
discrepancy between the two distributions indicates that our errors
are probably overestimated.
We believe this can be explained by the fact that only the brightest
RAVE stars have trigonometric parallaxes in the Hipparcos
catalogue. These brighter stars have higher S2N than the average RAVE
stars and so the true uncertainties on the stellar parameters are
actually smaller than our adopted values. The average S2N for these
624 stars is $\sim$64, which is twice the typical S2N ratio for RAVE
stars; correspondingly the uncertainties on the stellar parameters
will be smaller by a factor of 1.3 \citep[section 4.2.4 of ][]{Zwitter2008}.

We can quantify the overestimation in our distance errors for these
stars. The $3\sigma$ clipped standard deviation of the observed
distribution is 22.1\% and that of the predicted distribution is
27.8\%. To give the predicted distribution the same spread as the
observed distribution would require us to decrease the distance errors from
our method for these stars by $\sim$35\%.
Note that the $3\sigma$ clipping of this distribution is necessary
since a small fraction of our distances are in significant
disagreement with Hipparcos. Of the 624 stars in this cross-matched
sample, there are 3 with distance overestimates of more than 50\%, however closer inspection shows they qualify to be RC stars (\S \ref{sec:method:rave}). One more star qualifies as RC star and has a distance overestimate of 40\%, and one star with a $\logg = 2.8$ has a distance overestimate of 20\%. The systematic overestimation for possible RC stars and RGB stars is in agreement with our findings in the next section.

In Fig. \ref{fig:hip_systematics} we show the distribution of $(d_\text{RAVE} -
d_\text{Hipparcos})/d_\text{Hipparcos}$ as a function of the 2MASS colour
$\jk$ and of the three main stellar parameters ($\T$, \MH{},
$\logg$). We see no clear systematic
trends at a level of more than $\sim15\%$ in any of the properties
shown here, which implies that our method is producing reliable
distances for main-sequence stars.

\subsubsection{\object{M67} giants}
\label{sec:method:M67}

\begin{figure}
	\centerline{\includegraphics[keepaspectratio=true,scale=0.8]{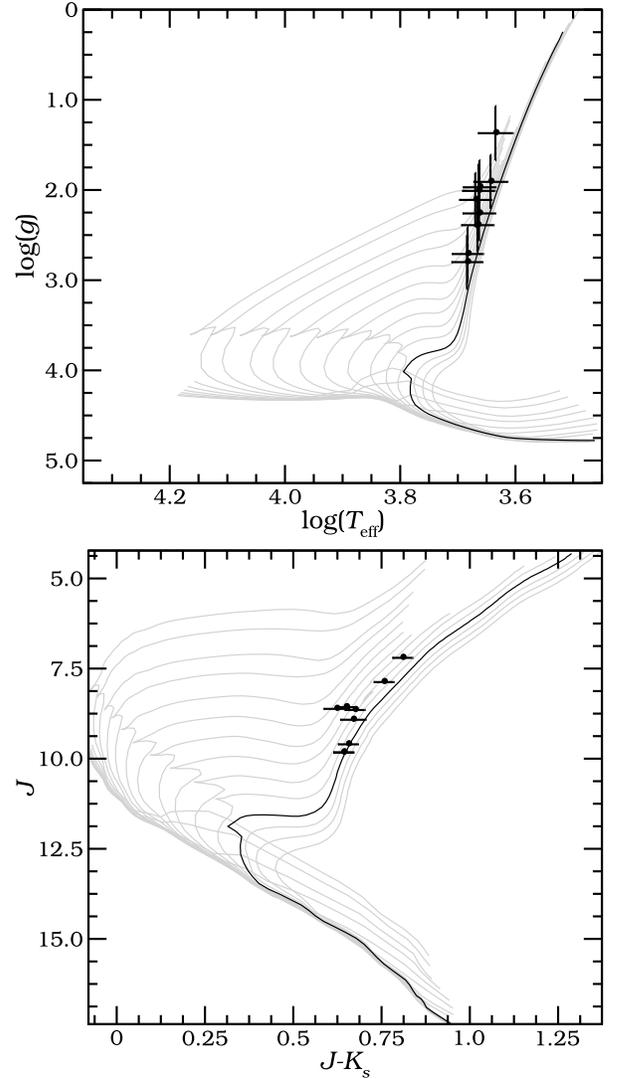}}
	\caption{{\bf Bottom}: CMD of M67 giants on top of
          theoretical solar-metallicity $Y^2$ isochrones, with the 4 Gyr isochrone in
          black. The isochrones are spaced logarithmically in age
          between 0.01 to 15 Gyr. Horizontal lines indicate 1$\sigma$
          uncertainties in $(J-K_s)$ and the uncertainties in the
          vertical direction are smaller than the size of the
          data-point. {\bf Top}: Similar to top panel, except now for $\logg$
          versus $\logT$. }\label{fig:m67}
\end{figure}

The results from the previous section give us confidence the method
works well for nearby main-sequence stars, but give us no indication of the
validity of the distances to giant stars.

Our preliminary RAVE dataset includes a small number of RGB
stars which are members of the old open cluster \object{M67}. As
the distance to M67 is relatively well known, this makes a perfect
test case for these stars. M67 has a distance modulus of
$(m-M)_V=9.70$, near-solar metallicities and an age of $\tau \approx
4$ Gyr \citep{VandenBerg2007}.

We identify members of M67 using the following criteria:
offset from the cluster centre of less than 0.55\degr;
heliocentric radial velocity within 3.3 \kms of the mean value of 32.3 \kms
\citep{2005A&A...438.1163K}, where this value of 3.3 \kms corresponds
to three times the uncertainty in the mean velocity;
signal to noise ratio S2N~$>$~20;
$\logg < 3.5$.
A total of 8 stars pass these criteria.
In Fig. \ref{fig:m67} we show these members, where one star is
observed twice. For these stars our method gives a distance of $\sim$1.82$\pm$0.27 kpc, more than twice the distance from the literature
\citep[$\sim0.8$ kpc;][]{VandenBerg2007}.
Note however that the 4 stars at $J \approx 8.8$ qualify as RC stars as defined in \S \ref{sec:method:rave}. If we exclude these stars then the distance to M67 is $1.48 \pm 0.36$ kpc. The distance estimate is now within 2 sigma of the assumed real distance of 0.8 kpc, but still systematically overestimated.
This overestimation can be understood when one considers the performance of
the stellar models. In the bottom panel we show the CMD of the members with
a set of isochrones for comparison. The black isochrone
is for an age similar to that of the M67 population ($4$ Gyr) and of solar
metallicity. At least one or both of the predicted colour and absolute magnitude of
the stars is incorrect. In the top panel we
show a plot of $\logg$ vs. $T_{\rm eff}$, which shows that the stars do not lie
on the isochrone in this plane either. Although the stars are within 1 or 2$\sigma$ from the $4$ Gyr isochrone, the deviation is systematic, particularly for the brighter RGB stars. This discrepancy will clearly
impair our method and hence it is not surprising that our distances
are affected. The difficulty of obtaining isochrones that match giants
is a long standing problem that is being addressed by various authors
\citep[e.g.][]{VandenBerg2008,Yadav2008}.

Therefore, given the limitations of the models used in this work,
our distances for stars with $\log(g) < 3$ should be
treated with caution. They can still be useful for analysing trends
in the data (\S \ref{sec:results}), but distances to
individual stars are likely to be inaccurate. Note as well that our simplification to treat RC as RGB stars will lead to an overestimation of their distance. We return to the issue
of stellar models in the discussion (\S \ref{sec:discussion:isochrones}).

\subsection{6D phase-space coordinates for stars in the RAVE dataset}
\label{sec:method:phasespace}

\begin{figure}
	\centerline{\includegraphics[keepaspectratio=true,scale=0.8]{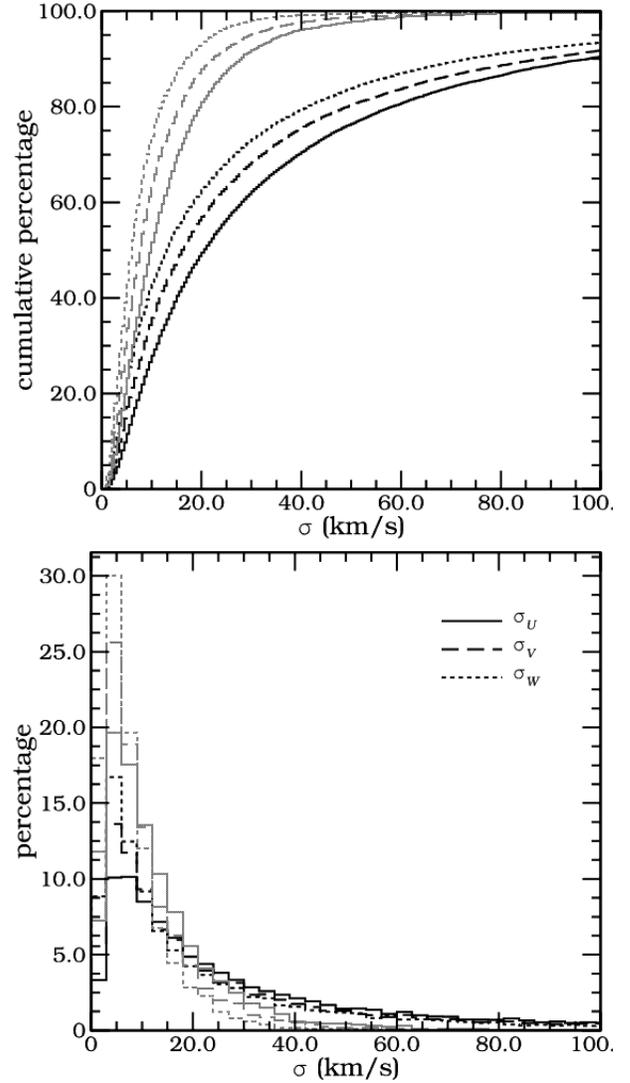}}
\caption{Distribution of uncertainties for velocity components
$U$ (solid line), $V$ (dashed line) and $W$ (dotted
line) velocities. This corresponds to the clean sample of \raveApply
stars (see \S\ref{sec:method:rave}). The black line includes all the
stars, while the grey line shows the distribution for main-sequence
stars (defined here as those with  $\logg{} > 4$).}
\label{fig:errors_histograms_vel}
\end{figure}

Besides providing distances to RAVE stars, we also provide full 6D
phase-space information derived using the radial velocities (from RAVE) and
the archival proper motions contained in the RAVE catalogue \citep[from the Starnet2, Tycho2, and UCAC2 catalogues; see][]{Zwitter2008}.

We use the Monte Carlo techniques described above to calculate 6D
phase-space coordinates assuming Gaussian errors on the observed
quantities (radial velocities, proper motions, etc). This is done using
 the transformations given by \citet{1987AJ.....93..864J}.

The coordinate system we use is a right-handed Cartesian coordinate
system centred on the Galactic Centre (GC): the $x$ axis is aligned
with the GC-Sun axis with the Sun located at $x=-8$ kpc; the $y$
axis pointing in the direction of rotation and the $z$ axis pointing
towards the Northern Galactic Pole (NGP). The velocities with respect
to the Sun in the directions of $(x,\,y,\,z)$ are ($U,\,V,\,W$) respectively,
with the rest frame taken at the Sun (such that the Sun is at
$(U_{\odot},\,V_{\odot},\,W_{\odot})=(0,\,0,\,0)$). Our final catalogue also
includes cylindrical polar coordinates $(v_\rho,\,v_\phi,\,v_z)$,
defined in a Galactic rest frame such that the local standard of rest (LSR) moves at $v_\phi = -220$ \kms. To transform from the rest frame of the Sun to the Galactic rest frame, we use $v_{\text{LSR}}$ = 220 \kms for the LSR and take the velocity of the Sun with respect to the LSR to be $(10.0\ \mathrm{\kms},\,5.25~\mathrm{\kms},\,7.17\ \mathrm{\kms})$ \citep{DehnenBinney1998}. A full description of the coordinate systems is given in Appendix \ref{sec:app_coords}.
An overview of the errors for $U$, $V$ and $W$ are shown in
Fig. \ref{fig:errors_histograms_vel}. We find that 7\,139 (44\% of the \raveApply) stars have errors less
than 20 \kms in all three velocity components, and $11\,742$ (73\%) have errors less
than 50 \kms. For the main-sequence stars this is 5\,425 (78\% of the \raveApplyMS) and 6\,832 (98\% of the \raveApplyMS) respectively.

\subsection{The catalogue}

Our catalogue is available for download from the webpage
{\tt http://www.astro.rug.nl/\~{}rave/} and is also hosted by the
CDS service VizieR.\footnote{\tt http://webviz.u-strasbg.fr} We aim to
update the catalogue as future RAVE data releases are issued. The
format of the catalogue is described in full in Appendix
\ref{sec:app_description}.

%%%%%%%%%%%%%%%%%%%%%%%%%%%%%%%%%%%%%%%%%%%%%%%%%%%%%%%%%%%%%%%%%%%%%%%
%%%%%%%%%%%%%%%%%%%%%%%%%%%%%%%%%%%%%%%%%%%%%%%%%%%%%%%%%%%%%%%%%%%%%%%
%%%%%%%%%%%%%%%%%%%%%%%%%%%%%%%%%%%%%%%%%%%%%%%%%%%%%%%%%%%%%%%%%%%%%%%
%%%%%%%%%%%%%%%%%%%%%%%%%%%%%%%%%%%%%%%%%%%%%%%%%%%%%%%%%%%%%%%%%%%%%%%
%%%%%%%%%%%%%%%%%%%%%%%%%%%%%%%%%%%%%%%%%%%%%%%%%%%%%%%%%%%%%%%%%%%%%%%
%%%%%%%%%%%%%%%%%%%%%%%%%%%%%%%%%%%%%%%%%%%%%%%%%%%%%%%%%%%%%%%%%%%%%%%
%%%%%%%%%%%%%%%%%%%%%%%%%%%%%%%%%%%%%%%%%%%%%%%%%%%%%%%%%%%%%%%%%%%%%%%
%%%%%%%%%%%%%%%%%%%%%%%%%%%%%%%%%%%%%%%%%%%%%%%%%%%%%%%%%%%%%%%%%%%%%%%
%%%%%%%%%%%%%%%%%%%%%%%%%%%%%%%%%%%%%%%%%%%%%%%%%%%%%%%%%%%%%%%%%%%%%%%
%%%%%%%%%%%%%%%%%%%%%%%%%%%%%%%%%%%%%%%%%%%%%%%%%%%%%%%%%%%%%%%%%%%%%%%
%%%%%%%%%%%%%%%%%%%%%%%%%%%%%%%%%%%%%%%%%%%%%%%%%%%%%%%%%%%%%%%%%%%%%%%
%%%%%%%%%%%%%%%%%%%%%%%%%%%%%%%%%%%%%%%%%%%%%%%%%%%%%%%%%%%%%%%%%%%%%%%

\section{Scientific Results}
\label{sec:results}

The main components of our Galaxy are the bulge, the halo and the
thin and thick disks. The thin disk has a scale height of $\sim$300
pc, while the thick disk scale height is $\sim$1 kpc \citep[e.g.][]{Juric2008}. The
disk is known to be dominated by metal rich stars, while halo
stars are in general metal poor (see \citet{Wyse2006} for a recent review).
To see if this is reflected in the RAVE data, we will now focus on how
the metallicity and kinematics change as a function of distance from
the plane.

\begin{figure*}
	\centerline{\includegraphics[keepaspectratio=true,scale=0.8]{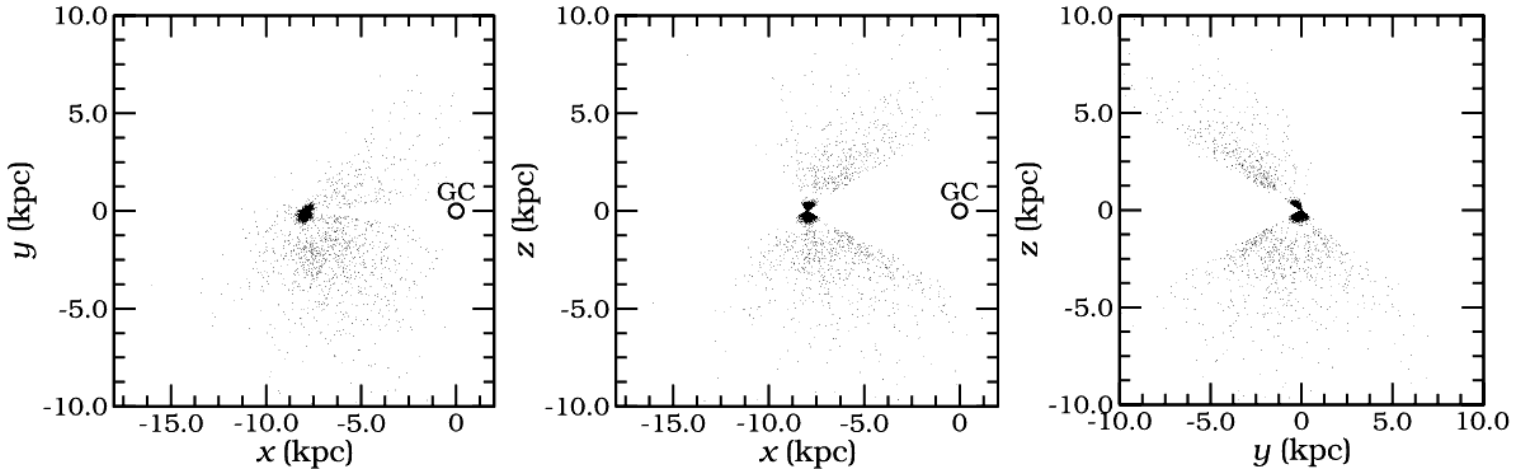}}
	\caption{
		The RAVE stars in galactic coordinates, the circle
                with label GC indicates the galactic centre (which we
                have assumed to be at a distance of 8 kpc from the Sun). We have
                only plotted those stars with distance error less than
                40\%.
	}\label{fig:rave_xyz2}
\end{figure*}

In Fig. \ref{fig:rave_xyz2} we show the spatial distribution of stars
in the RAVE dataset, where we have restricted ourselves to stars with
errors of less than 40\% in distance. As expected, we see a strong
concentration of stars within 1 kpc, illustrating that most of our
stars are nearby disk dwarfs. However, there  are also a number of stars
at much larger distances, which are giants probing into the Galactic
halo (although one should bear in mind that our giant distances are
likely to be unreliable; see \S \ref{sec:method:M67}).

\begin{figure}
	\centerline{\includegraphics[keepaspectratio=true,scale=0.8]{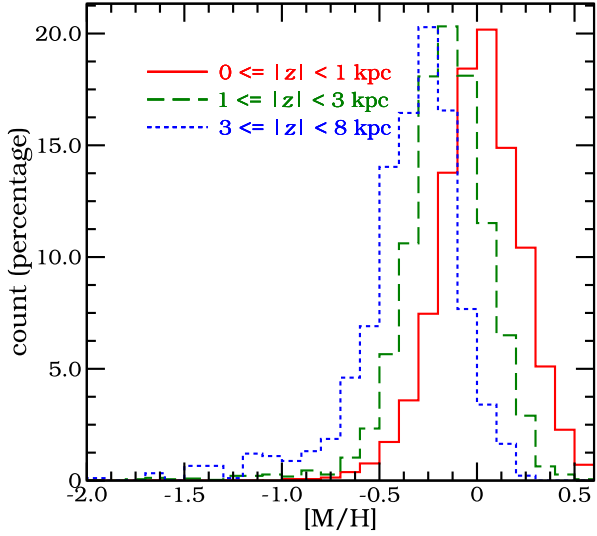}}
	\caption{
		Normalised metallicity distribution for
                stars in different bins of height above the Galactic
                plane, where we are only showing stars with distance
                error less than 75\%. As expected, stars further away
                from the Galactic plane are more metal
                poor.}\label{fig:rave_metallicity}
\end{figure}

Given this large span of distances, we can investigate the change in
metallicity as we move out of the Galactic plane. Since we
still have stars with non-negligible errors in distance, this analysis
will be subject to contamination from stars at different distances, so
we show only the relevant trends in our data. The resulting
distribution of metallicity for three $|z|$ bins is shown in
Fig. \ref{fig:rave_metallicity} for stars with relative distance error less than 75\%. It is clear that most of the stars in
the $|z|<1$ kpc bin are consistent with a solar-metallicity thin-disk
population, but as we move away from the plane the mean metallicity
decreases. In particular, a tail of metal-poor stars is evident for
$|z| > 3$ kpc, consistent with a halo population. The trends that we
are seeing are similar to those seen by \citet{Ivezic2008}, where the
metal-poor halo becomes apparent at \MH{} $\la -1$ for $|z| \ga 2$
kpc.

We now analyse the velocities of stars in our sample, restricting ourselves to
a high-quality subset of $\raveVel$ stars. For this sample we only use
those stars with distance error less than 40\%, proper motion error less
than 5 ${\rm mas}\,{\rm yr}^{-1}$ (in both components) and radial
velocity error less than 5 \kms.

\begin{figure}
	\centerline{\includegraphics[keepaspectratio=true,scale=0.8]{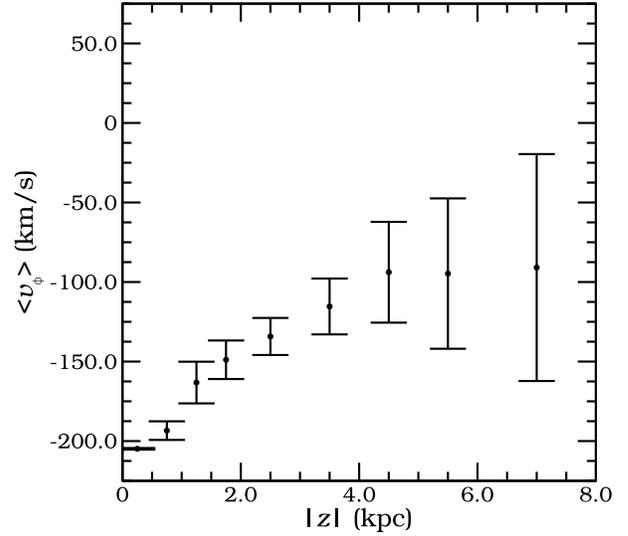}}
	\caption{
		Rotational velocity as a function of $|z|$ for the
                high-quality subset of \raveVel stars (see
                \S\ref{sec:results}). The
                error bars indicate $1\sigma$ uncertainty in the
                means. Note that the LSR has been assumed to move with $v_\phi = -220$
                \kms.}\label{fig:rave_velocities}
\end{figure}

In Fig. \ref{fig:rave_velocities} we have plotted the average $v_\phi$
(where $-220$~\kms corresponds to the LSR) in different bins of
$|z|$. It shows a decreasing rotational velocity as we move away from
the Galactic plane, which can be explained by a transition from a fast
rotating disk component, to a non-rotating (or slowly-rotating) halo.
As before, owing to our uncertainties in the giant distances, this
plot should only be used to draw qualitative conclusions.

\begin{figure*}
	\centerline{\includegraphics[keepaspectratio=true,scale=0.9]{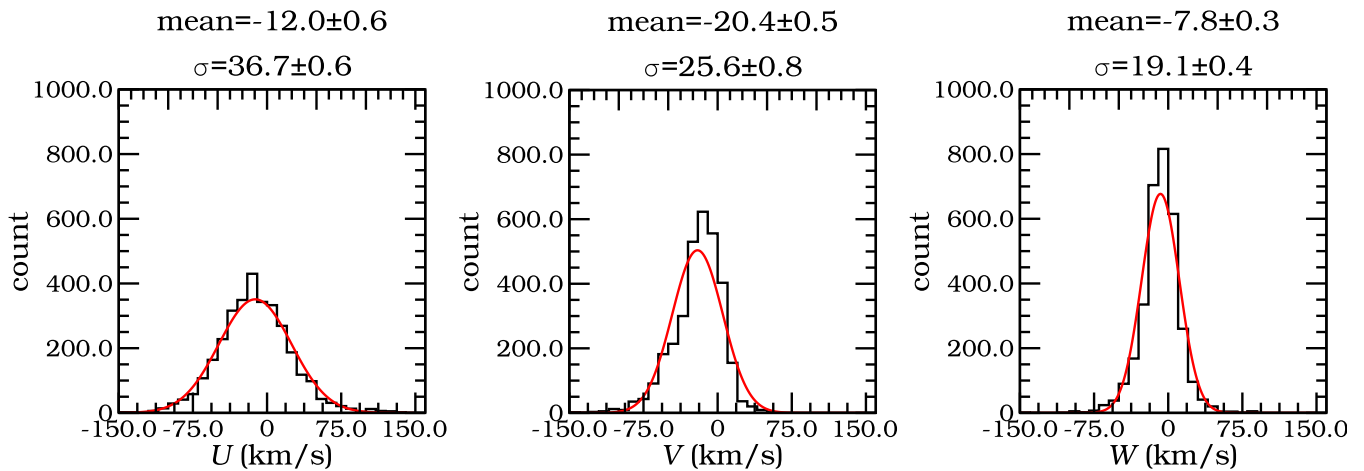}}
	\caption{
		Velocity distributions for the $U$, $V$ and $W$ components
                (histogram) and the best fit Gaussian (solid line) for
                high-quality volume-limited sample of \raveVelVol
                stars (see \S\ref{sec:results}).
		The velocity distributions for $U$ and $W$ are symmetric,
                showing a slight negative mean $U$ and $W$ owing to the
                solar motion with respect to the LSR. As expected, the $V$
                component shows an slight asymmetry, having a longer
                tail towards the slower rotating stars.}\label{fig:rave_veldist}
\end{figure*}

For nearby dwarfs ($\logg{} > 4$) the errors in velocity are relatively small,
therefore we refine our sample further by considering a volume-limited
sample. We use a cylindrical volume centred on the Sun with a radius
of 500 pc and a height of 600 pc (300 above and below the Galactic
plane). This sample, which contains \raveVelVol stars, has average
errors of $(8.2~\text{\kms},\,6.3~\text{\kms},\,5.1~\text{\kms})$ in the $(U,\,V,\,W)$ directions, respectively.
The velocity distributions for these
stars are shown in Fig. \ref{fig:rave_veldist} and the corresponding means
and velocity dispersions are given in Table \ref{tab:vel}. The uncertainties are
obtained by a bootstrap method. Note that these distributions will be
broadened by the observational errors, but we have not taken this into
account when calculating these variances.
For this sample, we also tabulate the full
velocity dispersion tensor $\sigma_{ij}$. As has been found by
previous studies \citep[e.g.][]{DehnenBinney1998}, the $\sigma_{UV}$
term is clearly non-zero ($\sigma^2_{UV} = 108.0 \pm 25.7~\textrm{km$^2$\,s$^{-2}$}$). For this component we can calculate the vertex deviation,
\begin{align}
l_v = \frac{1}{2} \arctan \left( 2\frac{\sigma_{UV}^2}{\sigma_{U}^2-\sigma_{V}^2}\right),
\end{align}
which is a measure of the orientation of the $UV$ velocity
ellipsoid. We find $l_v = 8.7 \pm 2.0^\circ$, which is comparable to
the value of $10^\circ$ found for stars with $(B-V)\ga 0.4$ in
the immediate solar neighbourhood by \citet{DehnenBinney1998}.
The uncertainties on the other two cross-terms ($\sigma_{UW}^2$ and
$\sigma_{VW}^2$) are too large to allow us to detect any weak
correlations that might be present.

\begin{figure*}

  \includegraphics[scale=0.9]{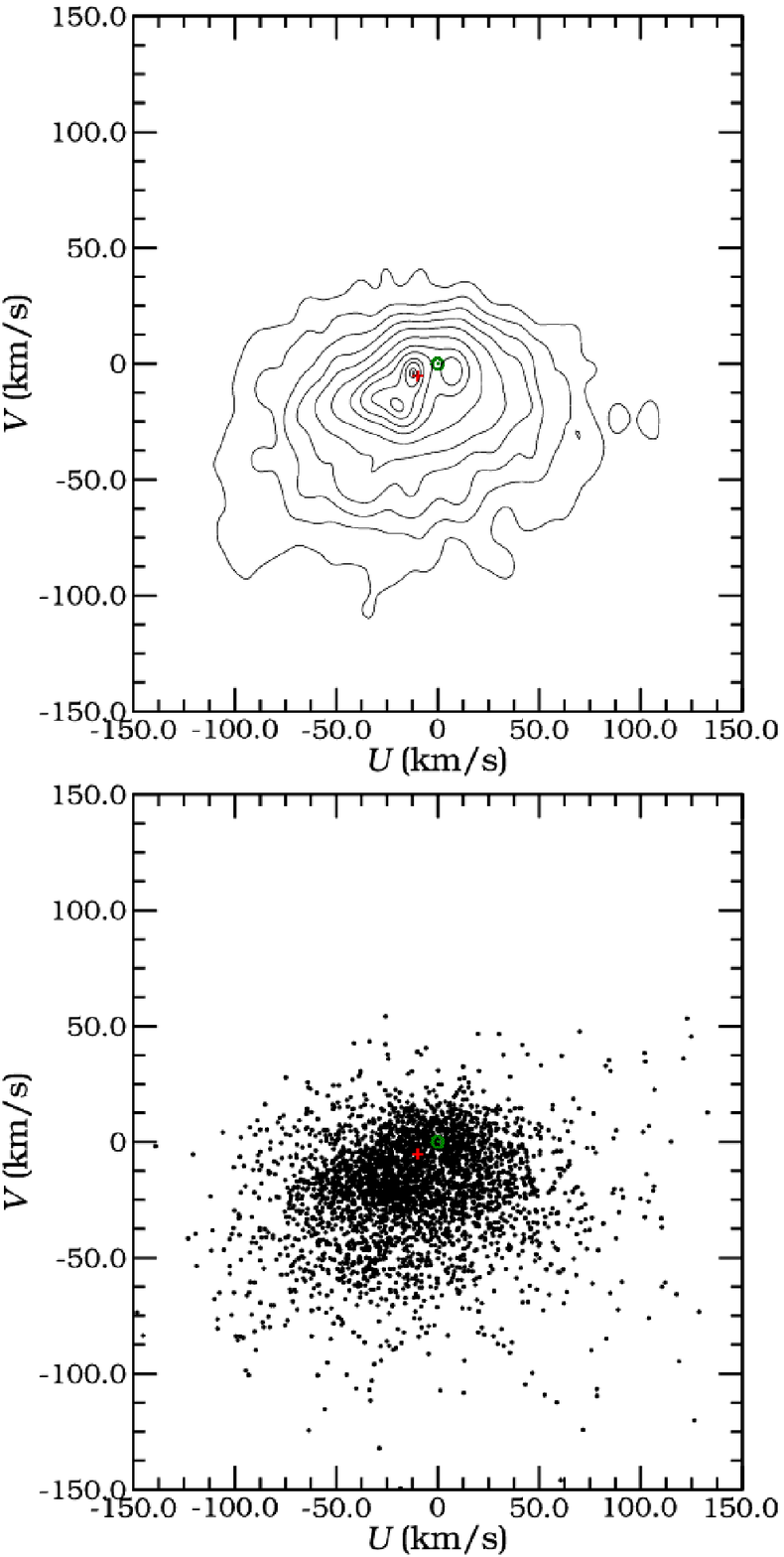}\hfil
  \includegraphics[scale=0.9]{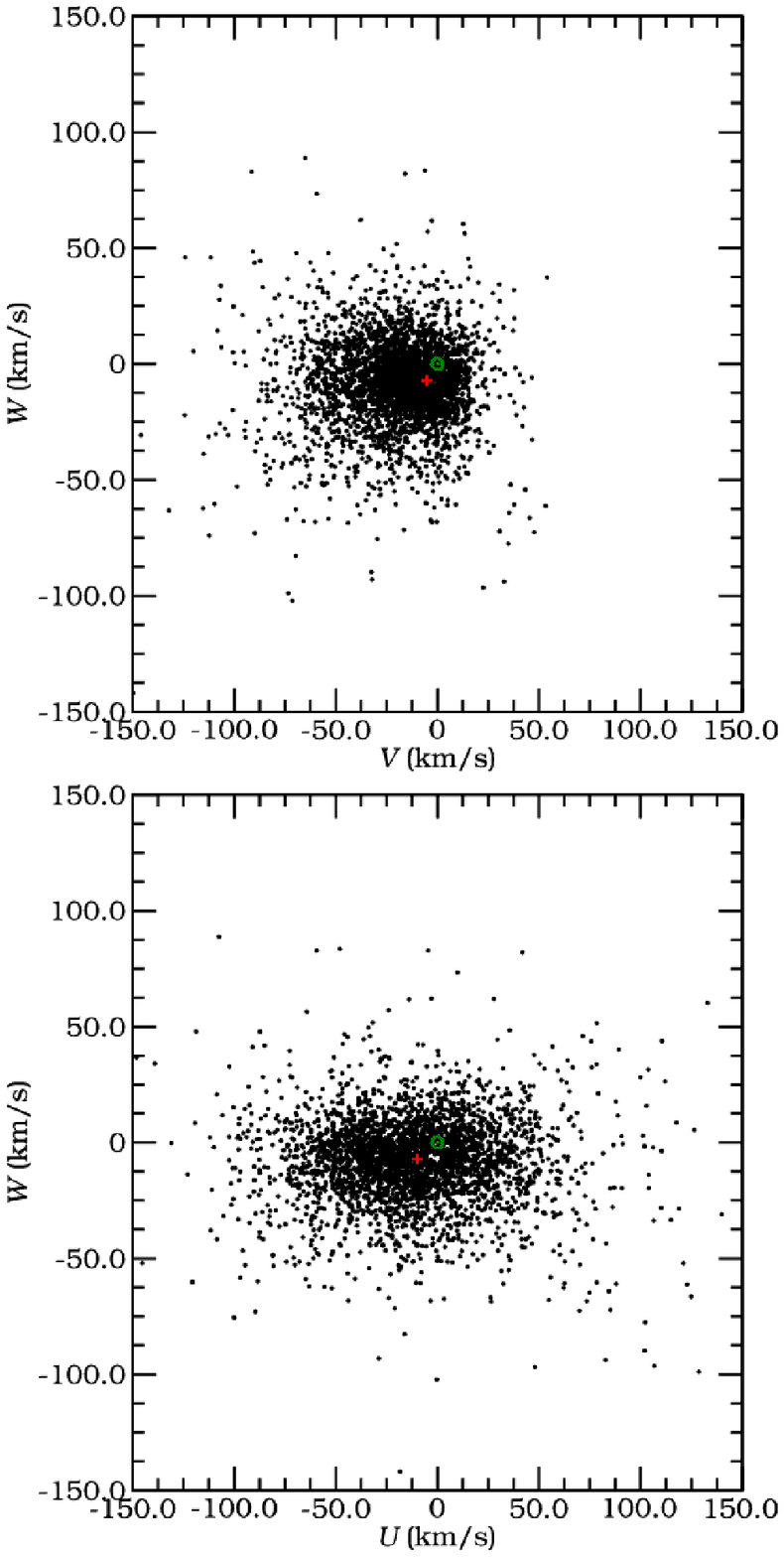}

	\caption{The $UV$, $UW$ and $VW$ velocity distributions for the
          high-quality volume-limited sample of \raveVelVol stars (see
          \S\ref{sec:results}). The upper-left panel shows isodensity
          contours for the $UV$ plane, where the contours contain 2, 6,
          12, 21, 33, 50, 68, 80, 90, 99 and 99.9 percent of the stars.
          The red $+$ symbol marks the LSR \citep{DehnenBinney1998}
          and the green $\odot$ symbol marks the solar velocity
          $(0,0)$.}\label{fig:rave_UV}
\end{figure*}

Close inspection of the middle panel of Fig. \ref{fig:rave_veldist}
shows an asymmetric distribution for the $V$ component, with a longer
tail towards lower velocities. This is due to two effects. The first
is that we are seeing the well-known asymmetric drift, where
populations of stars with larger velocity dispersions lag behind the
LSR \citep{BM}.
Secondly, it is known that the velocity distribution of the
solar neighbourhood is not smooth \citep[see, e.g.][]{Chereul1998,Dehnen1998,Nordstrom2004}. This
issue is further illustrated in Fig. \ref{fig:rave_UV}, where we show
the distribution of velocities in the $UV$-plane. A slight over-density
of stars around $U \approx -50$~\kms, $V \approx -50$~\kms can be seen which
will affect the symmetry of the $V$ velocity component. This
over-density is called the Hercules stream, and is thought to be
due to a resonance with the bar of our Galaxy \citep{Dehnen2000,Fux2001}.

It should be noted that all velocities are with respect to the Sun,
which implies that the Sun's $U$ and $W$ velocity with respect to the
LSR are the negative of the mean $U$ and $W$ in our sample.
Due to the asymmetric drift, the $V$ velocity of the complete sample
of stars is not equal to the negative of the $V$ velocity of Sun with
respect to the LSR \citep{BM}. The velocities and dispersions are in
reasonable agreement with the results of \citet{Famaey2005} and
\citet{DehnenBinney1998} even though we are using different samples
from those examined in these previous studies (e.g. probing different volumes or
types of stars).

\begin{table}
 \begin{tabular}{cccc}

Mean & $\overline{U}$ & $\overline{V}$ & $\overline{W}$\\\hline
(\kms) & $-12.0 \pm 0.6$ & $-20.4 \pm 0.5$ & $-7.8 \pm 0.3$\\\\

Standard Deviation & $\sigma_U$ & $\sigma_V$ & $\sigma_W$\\\hline
(\kms) & $36.7 \pm 0.6$ & $25.6 \pm 0.8$ & $19.1 \pm 0.4$\\\\

Covariance & $\sigma^2_{UV}$ & $\sigma^2_{UW}$ & $\sigma^2_{VW}$\\\hline
$(\textrm{km$^2$\,s$^{-2}$})$ & $108.0 \pm 25.7$ & $-19.7 \pm 17.3$ & $12.8 \pm 16.2$

 \end{tabular}
\caption{Means, standard deviations and covariances for $U$, $V$
  and $W$ velocities corresponding to the high-quality volume
  limited sample of \raveVelVol stars (see \S\ref{sec:results}).
}
\label{tab:vel}
\end{table}

\begin{comment}
\begin{table}
 \begin{tabular}{cccc}

Mean & $\overline{U}$ & $\overline{V}$ & $\overline{W}$\\\hline
(\kms) & $-11.5 \pm 0.6$ & $-19.4 \pm 0.4$ & $-7.5 \pm 0.3$\\\\

Standard Deviation & $\sigma_U$ & $\sigma_V$ & $\sigma_W$\\\hline
(\kms) & $34.9 \pm 0.6$ & $24.3 \pm 0.7$ & $18.3 \pm 0.3$\\\\

Covariance & $\sigma^2_{UV}$ & $\sigma^2_{UW}$ & $\sigma^2_{VW}$\\\hline
$(\textrm{km$^2$\,s$^{-2}$})$ & $97.1 \pm 21.3$ & $2.5 \pm 15.7$ & $4.6 \pm 14.0$

 \end{tabular}
\caption{OLD: Means, standard deviations and covariances for $U$, $V$
  and $W$ velocities. This corresponds to the high-quality volume
  limited sample of \raveVelVol stars (see \S\ref{sec:results}).
}
\label{tab:velold}
\end{table}
\end{comment}

\section{Discussion}
\label{sec:discussion}

\subsection{The influence of the choice of stellar models}
\label{sec:discussion:isochrones}

\begin{figure*}
	\centerline{\includegraphics[keepaspectratio=true,scale=0.9]{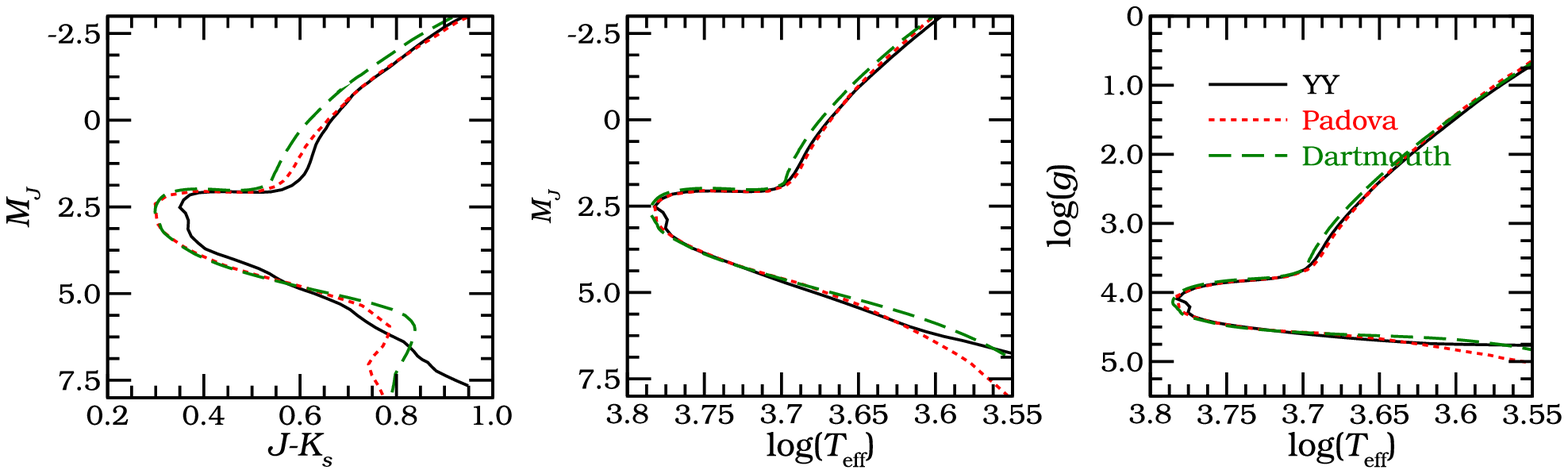}}
	\caption{A comparison of isochrones from three separate
          groups: Yale-Yonsei (black), Padova (red), Dartmouth
          (green). We have chosen isochrones with age 5 Gyr, Z =
          0.019, \enh{} = 0.}\label{fig:isochrone_compare}
\end{figure*}

The method described in \S \ref{sec:method:method} clearly relies on
the ability of stellar models to accurately predict the observed
parameters. Therefore it is worth briefly discussing the potential
difficulties which may arise from this assumption.

As was discussed in \S\ref{sec:method:method} we have chosen to use
the Yale-Yonsei ($Y^2$) models \citep{Demarque2004}, but there are
several groups who make stellar models. In
Fig.~\ref{fig:isochrone_compare} we compare isochrones (with age 5
Gyr, $\metal=0.019$, $\enh=0$) from the following three groups:
the $Y^2$ group \citep{Demarque2004}, the Padova group
\citep{Marigo2008} and the Dartmouth group \citep{Dotter2008}.
The latter paper can be consulted for a more detailed
comparison of the various groups' theoretical models \citep[see
also][]{Glatt2008}.

In general the three curves in the $\logT$-$\logg$ plane and $\logT$-$M_J$
plane show reasonably good agreement, certainly within the
observational errors of the RAVE data (see \S \ref{sec:rave:data}). The
largest discrepancy is for the cool dwarfs ($\logT <  3.65$), but we do
not believe this should have any significant effect on our results as
we have very few stars in this regime.
When one considers the $\jk$-$M_J$ plane the situation is less
satisfactory, probably due to the $\T$-colour transformations.

To assess whether our decision to use the $Y^2$ models has any
serious effect on our results, we repeat the analysis presented in
Section \ref{sec:app:testing} using the Dartmouth models. We find that
this has very little influence; there is no noticeable improvement for
either the Hipparcos dwarfs or the M67 giants. Therefore we conclude
that our method is not particularly sensitive to the choice of stellar
models. However, one should still  bear in mind that, by
definition, our method will be limited by any problems or deficiencies
in the adopted set of isochrones.

\subsection{Comparisons to other work}
\begin{figure}
	\centerline{\includegraphics[keepaspectratio=true,scale=0.9]{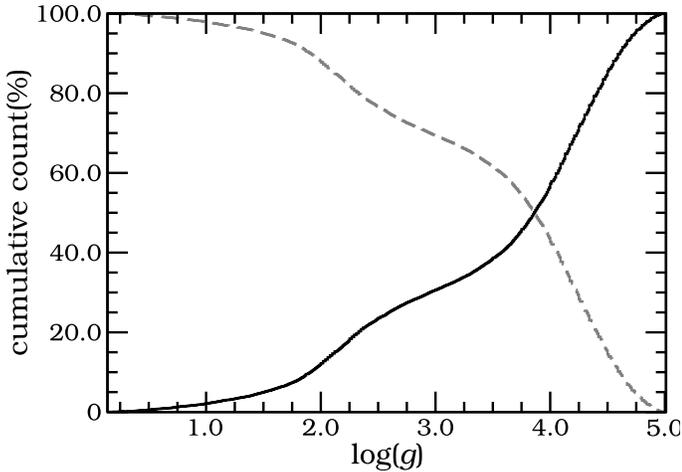}}
	\caption{Cumulative distribution of $\logg$, the black line shows the \% of stars below a certain $\logg$, while the dashed grey line shows stars above a certain $\logg$.
	}\label{fig:rave_klementcompare}
\end{figure}

\citet[][ hereafter K08]{Klement+2008} have used a different method to
derive distances for RAVE stars, seemingly obtaining significantly
smaller errors than ours. They calibrated a photometric distance
relation (relating $V_T-H$ to $M_V$) using stars from Hipparcos catalogue with
accurate trigonometric parallaxes, combined with photometry from
Tycho-2, USNO-B and 2MASS. This method was then applied to the first
RAVE data release \citep[DR1;][]{Steinmetz2006}.
Although the number of stars analysed by K08 is
similar to that considered here ($\sim25\,000$), they obtain $\sim7\,000$
stars with distance errors smaller than 25\%, while we have only
431 stars with distance errors smaller than 25\%.

The K08 method relies on stars being on the main sequence. However,
from the values of \logg we can now show that of order half the RAVE
stars are giants: in Fig. \ref{fig:rave_klementcompare} we show the
cumulative distribution of $\logg$, showing that main-sequence stars
($\logg > 4$, see also Fig. \ref{fig:colorisochrones}) are only
$\sim 40$\% of the whole sample.
Therefore it is clear that a
large fraction of the RAVE sample are giants, subgiants or close to
the main-sequence turn-off.
This will undoubtedly affect the results presented in
K08. For example, their plot of the $UW$ velocity
distribution is evidently suffering from significant systematics as can
be seen from the correlation between the $U$ and $W$ velocities. Previous
studies of local samples of stars have not found such a
correlation.
Our distribution of $UW$ shows no such strong correlation
(Fig. \ref{fig:rave_UV}) and $\sigma_{UW}$ is consistent with 0.
Even in samples of stars out of the plane where one
might expect correlations to appear, there is no evidence for such a
pronounced level of correlation \citep{Siebert+2008}.

As well as the problem of misclassified giant stars, additional
factors that will adversely affect the K08 distances include: the
metallicity distribution of the local RAVE sample will probably differ
from that of Hipparcos due to the fact that RAVE probes a different
magnitude range (and hence volume); or that K08 use the $V$-band which
is more prone to reddening than our choice of $\jk$. With regard to
this latter point, we can repeat the simple analysis presented in
Section \ref{sec:method:method}. For a typical star 250 pc away, given
the RAVE field-of-view the dust maps of \citet{Sch1998} predict
extinction of $\sim0.1$ mag in $V$ (with corresponding distance error
of $\sim5\%$) and reddening of $\sim0.1$ mag in $(V-H)$.

\section{Conclusion}
\label{sec:conc}
We have presented a method to derive absolute magnitudes, and
therefore distances, for RAVE stars using stellar models.
It is based on the use of stellar model fitting in metallicity,
$\logg$, $\T$ and colour space.

We find that our method reliably estimates distances for main-sequence
stars, but there is an indication of potential systematic problems
with giant stars owing to issues with the underlying stellar models.
The uncertainties in the estimated absolute magnitudes for RGB stars are
found to depend mainly on the uncertainties in $\logg$, while for main-sequence
stars the accuracy of $\T$ is also important (\S
\ref{sec:method:test}). For the RAVE data the uncertainties in $\logg$
and $\T$ give rise to relative distance uncertainties in the range 30\%-50\%,
although from cross-matching with Hipparcos (\S
\ref{sec:app:testing:hipp}) it appears that our uncertainties may be
overestimated for the brighter stars (with higher signal-to-noise spectra).
It is important to note that that some 10\% of the RAVE stars may
be on the red clump, but these are treated as RGB by our pipeline,
and hence their distances may be systematically biased.

As can be seen in the results section (\S \ref{sec:results}), the data
accurately reflect the known properties of halo and disk stars of
the Milky Way. A variation in metallicity and $v_\phi$ was found away
from the Galactic plane, corresponding to an increase in the fraction
of metal-poor halo stars. Existing substructure in the $UV$ velocity
plane was recovered, as was the vertex deviation. Upon completion
the RAVE survey will have observed a factor of up to $\sim$20 times more
stars than analysed here. Clearly this will be a hugely valuable
resource for studies of the Galaxy.

In future the Gaia satellite mission \citep{Perryman2001}
will revolutionise this field, recording distances to millions of
stars with unprecedented accuracy. However, for large numbers of Gaia
stars it will not be possible to accurately constrain the distance due
to them being too far away or too faint, which implies that it is crucial to
develop techniques such as ours for reliably estimating distances.

In the near term it will be possible to improve the accuracy of our
pipeline by calibrating it through observations of clusters; a
technique which has been used with great success by the Sloan Digital Sky
Survey \citep{Ivezic2008}. Within the RAVE collaboration a project is
underway to obtain data for cluster stars \citep[e.g.][]{Kiss2007} and
we aim to incorporate this into future analyses. This may allow us to
reduce or remove the reliance on stellar models, which will lessen one
of the major sources of uncertainty in our work.
Our pipeline will allow us to fully utilise current surveys such as
RAVE, and also places us in an ideal position exploit future
large-scale spectroscopic surveys that will be enabled by upcoming
instruments such as LAMOST.

\begin{acknowledgements}

We thank the referee for useful suggestions that helped
improve the paper. We also thank Heather L. Morrison and Michelle L. Wilson for their helpful suggestions. M.A.B and A.H. gratefully acknowledge the the Netherlands Research School for Astronomy (NOVA) for
financial support. M.C.S. and A.H. acknowledge financial support from the Netherlands
Organisation for Scientific Research (NWO).
M.C.S. acknowledges support from the STFC-funded `Galaxy Formation and
Evolution' program at the Institute of Astronomy, University of
Cambridge.

Funding for RAVE has been provided by the Anglo-Australian
Observatory, by the Astrophysical Institute Potsdam, by the Australian
Research Council, by the German Research foundation, by the National
Institute for Astrophysics at Padova, by The Johns Hopkins University,
by the Netherlands Research School for Astronomy, by the Natural
Sciences and Engineering Research Council of Canada, by the Slovenian
Research Agency, by the Swiss National Science Foundation, by the
National Science Foundation of the USA (AST-0508996), by the
Netherlands Organisation for Scientific Research, by the Science and
Technology Facilities Council of the UK, by Opticon, by
Strasbourg Observatory, and by the Universities of Basel, Cambridge,
Groningen and Heidelberg.

The RAVE web site is at www.rave-survey.org.

\end{acknowledgements}

\bibliographystyle{aa}
\bibliography{12471man}

\appendix
\section{Description of RAVE catalogue with phase-space coordinates}
\label{sec:app_description}
We present the results of our distance determinations and
corresponding phase-space coordinates as a comma separated values
(CSV) file, with headers.
The columns are described in Table \ref{tab:cat}. See
\citet{Steinmetz2006,Zwitter2008} for a more detailed description of the
RAVE data.

\begin{table*}[!htp]
\caption{A full description of the catalogue.}
\label{tab:cat}
\begin{tabular}{llll}
\hline
Field name & Units & Type & Description \\
\hline\hline
OBJECT\_ID & ~ & string & RAVE internal identifier\\
RA & deg & float & Right ascension (J2000)\\
DE & deg & float & Declination (J2000)\\
Glon & deg & float & Galactic longitude\\
Glat & deg & float & Galactic latitude\\
RV & \kms & float & Weighted mean of available radial velocities\\
eRV & \kms & float & Weighted error of available radial velocities\\
pmRA & mas\,yr$^{-1}$ & float & Proper motion RA\\
pmDE & mas\,yr$^{-1}$ & float & Proper motion DE\\
epmRA & mas\,yr$^{-1}$ & float & Error proper motion RA\\
epmDE & mas\,yr$^{-1}$ & float & Error proper motion DE\\
Teff & Kelvin & float & Arithmetic mean of available temperatures \\
nTeff & & int & Number of observations having \T{}\\
logg & $\log(\frac{cm}{s^2})$ & float & Arithmetic mean of available surface gravities\\
nlogg & & int & Number of observations having \logg\\
MH & dex & float & Arithmetic mean of RAVE uncalibrated metallicity (\mH) \\
 &  & & abundance\\
nMH & & int & Number of observations having \mH\\
MHcalib & dex & float & Arithmetic mean of RAVE calibrated metallicity (\MH)\\
&  &  & abundance\\
nMHcalib & & int & Number of observations having \MH\\
AM & dex & float & Arithmetic mean of RAVE alpha enhancement (\enh)\\
nAM & & int & Number of observations having \enh\\
Jmag & mag & float & 2MASS $J$ magnitude \\
eJmag & mag & float & error on Jmag \\
Kmag & mag & float & 2MASS $K_s$ magnitude \\
eKmag & mag & float & error on Kmag \\
Mj & mag & float & Absolute magnitude in $J$ band (from fitting method) \\
eMj & mag & float & Error on $M_J$ \\
distance & kpc & float & Distance from $M_J$ and $J$ \\
edistance & kpc & float & Error on distance \\
xGal & kpc & float & Galactic $x$ coordinate\footnotemark[2] \\
exGal & kpc & float & Error on $x$ \\
yGal & kpc & float & Galactic $y$ coordinate\footnotemark[2] \\
eyGal & kpc & float & Error on $y$ \\
zGal & kpc & float & Galactic $z$ coordinate\footnotemark[2] \\
ezGal & kpc & float & Error on $z$ \\
U & \kms & float & Galactic velocity on $x'$ direction w.r.t the Sun ($U$)\footnotemark[2] \\
eU & \kms & float & Error on $U$ \\
V & \kms & float & Galactic velocity on $y'$ direction w.r.t the Sun ($V$)\footnotemark[2] \\
eV & \kms & float & Error on $V$ \\
W & \kms & float & Galactic velocity on $z'$ direction w.r.t the Sun ($W$)\footnotemark[2] \\
eW & \kms & float & Error on $W$ \\
vxGal & \kms & float & Galactic velocity on $x$ direction in Galactic rest frame ($v_x$)\footnotemark[2] \\
evxGal & \kms & float & Error on $v_x$ \\
vyGal & \kms & float & Galactic velocity on $y$ direction in Galactic rest frame ($v_y$)\footnotemark[2] \\
evyGal & \kms & float & Error on $v_y$ \\
vzGal & \kms & float & Galactic velocity on $z$ direction in Galactic rest frame ($v_z$)\footnotemark[2] \\
evzGal & \kms & float & Error on $v_z$ \\
Vr & \kms & float & Galactic velocity on $\rho$ direction in Galactic rest frame ($v_\rho$)\footnotemark[2] \\
eVr & \kms & float & Error on $v_\rho$ \\
Vphi & \kms & float & Galactic velocity on $\phi$ direction in Galactic rest frame ($v_\phi$)\footnotemark[2] \\
eVphi & \kms & float & Error on $v_\phi$ \\

\hline
\end{tabular}
\end{table*}

\section{Coordinate systems}
\label{sec:app_coords}

\begin{figure}
\centerline{\includegraphics[keepaspectratio=true,scale=0.7]{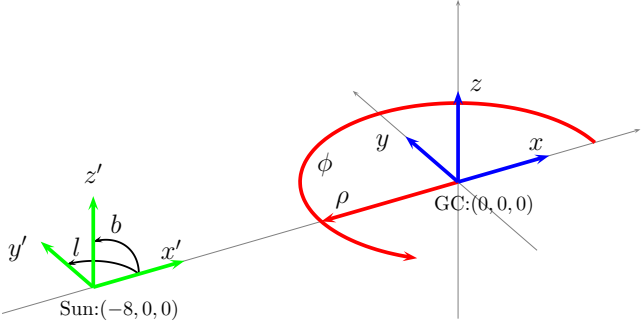}}
	\caption{Overview of the Galactic coordinates. The Sun is found at $(x,y,z)=(-8,0,0)$. $l$ and $b$ are the Galactic sky coordinates.} \label{fig:coord_pos}
\end{figure}

\begin{figure}
\centerline{\includegraphics[keepaspectratio=true,scale=0.7]{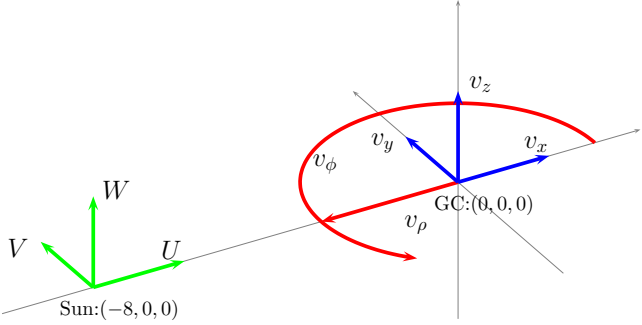}}
	\caption{Overview of Galactic coordinate systems. $U,V,W$ velocities are with respect to the Sun and are aligned with the $x',y',z'$ coordinate system. $v_x,v_y,v_z$ are Cartesian velocities, and $v_\rho,v_\phi$ are cylindrical velocities, both with respect to the Galactic rest frame.}\label{fig:coord_vel}
\end{figure}

The $x',y',z'$ coordinate system we use is a right handed Cartesian coordinate system centred on the Sun indicating positions, with the $x'$ axis pointing from the Sun to the Galactic Centre (GC), the $y'$ axis pointing in the direction of rotation and the $z'$ axis pointing towards the Northern Galactic Pole (NGP). The $x,y,z$ coordinate system is similar to the $x',y',z'$ coordinate system, but centred on the GC, assuming the Sun is at $(x,\,y,\,z)=(-8~\text{kpc},\,0,\,0)$. An overview can be found in Fig. \ref{fig:coord_pos} with Galactic longitude $(l)$ and latitude $(b)$ shown for completeness.

The velocities with respect to the Sun in the directions of $x,\,y,\,z$ are $U,\,V,\,W$ respectively. For velocities of nearby stars, a Cartesian coordinate system will be sufficient, but for large distances, a cylindrical coordinate system makes more sense for disk stars. To calculate these coordinates, we first have to transform the $U,\,V,\,W$ velocities to the Galactic rest frame, indicated by $v_x,\,v_y,\,v_z$ as shown in Fig. \ref{fig:coord_vel}.
Assuming a local standard of rest (LSR) of $v_{\text{LSR}}$~=~220~\kms, and the velocity of the Sun with respect to the LSR from \citet{DehnenBinney1998}, we find:
\begin{align}
v_x &= U + 10.0\ \mathrm{\kms},\\
v_y &= V + v_{\text{LSR}} + 5.25~\mathrm{\kms},\\
v_z &= W + 7.17\ \mathrm{\kms}.
\end{align}

The relations between Cartesian ($x,\,y,\,z$) and cylindrical coordinates ($\rho,\,\phi,\,z)$ are:
\begin{align}
x &= \rho \cos(\phi),\\
y &= \rho \sin(\phi),\\
z &= z,\\
\rho^2 &= x^2+y^2,\\
\tan(\phi) &= \frac{y}{x}.
\end{align}

We can use this to find the velocities in the directions of $\rho$ and $\phi$:
\begin{align}
v_\rho &= \frac{d\rho}{dt} = \frac{x v_x+y v_y}{\rho},\\
v_\phi &= \rho \frac{d\phi}{dt} = \frac{x v_y-y v_x}{\rho}.
\end{align}

Note that the direction of $\phi$ is anti-clockwise, meaning that the LSR is at $(v_\rho,\,v_\phi,\,v_z) = (0~\text{\kms},\,-220~\text{\kms},\,0~\text{\kms})$.
\\

\newpage

\footnotetext[2]{See \S \ref{sec:method:phasespace} for a description.}

\end{document}